\documentclass[notitlepage,twocolumn,prl,longbibliography]{revtex4-1}
\usepackage{amsmath}
\usepackage{changes}
\usepackage{graphicx}
\usepackage[space]{grffile} 
\usepackage{gensymb}
\usepackage{bm}
\usepackage{float}
\usepackage{amssymb,amsfonts,amsmath,amsbsy,bm,t1enc,latexsym}
\usepackage{changes}
\usepackage{soul}
\usepackage[super]{nth}
\usepackage[normalem]{ulem}
\usepackage[
bookmarks=true,
colorlinks=true,
linkcolor=blue,
urlcolor=blue,
citecolor=blue,
pdftex,
bookmarks=true,
linktocpage=true, 
hyperindex=true
]{hyperref}
\usepackage{hyperref}
\usepackage{caption}
\begin{document}	
	
\title{Dual chirped micro-comb based parallel ranging at megapixel-line rates}
	
	\author{Anton Lukashchuk}
	\email{anton.lukashchuk@epfl.ch}
	\author{Johann Riemensberger}
	\email{johann.riemensberger@epfl.ch}
	\author{Maxim Karpov}
    \author{Junqiu Liu}
	\author{Tobias J. Kippenberg}
	\email{tobias.kippenberg@epfl.ch}
	\affiliation{Laboratory of Photonics and Quantum Measurements (LPQM), Swiss Federal Institute of Technology Lausanne (EPFL), CH-1015 Lausanne, Switzerland}
	
	\date{\today}
	
	
	\maketitle



\textbf{
Laser based ranging (LiDAR) - already ubiquitously used in industrial monitoring, atmospheric dynamics, or geodesy - is key sensor technology. 
Coherent laser ranging \cite{Bostick1967,Uttam1985, Piggott2020}, in contrast to time-of-flight approaches, 
is immune to ambient light, operates continuous wave allowing higher average powers, 
 and yields simultaneous velocity and distance information. 
State-of-the-art coherent single laser-detector architectures reach hundreds of kilopixel per second rates \cite{Okano2020}. While emerging applications such as autonomous driving, robotics, and augmented reality mandate megapixel per second point sampling to support real-time video-rate imaging.
Yet, such 
rates of coherent LiDAR have not been demonstrated.
Here we report a swept dual-soliton microcomb technique enabling coherent ranging  and velocimetry at megapixel per second line scan measurement rates with up to 64 spectrally dispersed optical channels.
 It is based on recent advances in photonic chip-based microcombs~\cite{Kippenberg2018, Riemensberger2020} that offer a solution to reduce complexity both on the transmitter and receiver sides. 
Multi-heterodyning two synchronously frequency-modulated microcombs yields distance and velocity information of all individual ranging channels on a single receiver alleviating the need for individual separation, detection, and digitization.
 The reported LiDAR implementation is hardware-efficient, compatible with photonic integration, and demonstrates the significant advantages of acquisition speed  afforded by the convergence of optical telecommunication and metrology technologies.
We anticipate our research will motivate further investigation of frequency swept microresonator dual-comb approach in the neighboring fields of linear and nonlinear spectroscopy, optical coherence tomography.}


Three dimensional (3D) imaging based on 
lasers (LiDAR) is ubiqitiously used in numerous applications, ranging from airborne imaging for cartography of geological sites~\cite{Canuto2018} and urban-planning, to satellite-based applications in space. 
In recent years, driven by the large investments and development under way in autonomous driving \cite{Urmson2008}, drone technology \cite{Almeida2019} and industrial inspection \cite{Breitbarth2021, Breitenmoser2012} there has been a surge of interest in more sophisticated laser ranging systems.
LiDAR allows to maintain excellent angular resolution at long range, works reliably in a variety of weather, illumination and target conditions that impede direct camera imaging. While most commercial implementations of LiDAR employ incoherent detection of the intensity of reflected light, coherent detection of the backreflected signal using a copy of the transmitted optical waveform 
is intrinsically resilient to crosstalk and interference from ambient sunlight detection \cite{Behroozpour2017}. Furthermore, it achieves high depth resolution dependent on chirp excursion without the need for high-bandwidth electronics \cite{Agishev2006,Rogers2020}, and, importantly, gives both distance and velocity via the Doppler effect for each pixel \cite{Pierrottet2008}, significantly facilitating classification. 
One challenge to harvesting the inherent advantages of FMCW LiDAR for 3D imaging is to overcome the frame-rate acquisition bottleneck that 
 is imposed by tunable diode laser sources that trade-off tunability versus linewidth \cite{Amann1992} and artificial Doppler broadening due to the mechanical tilt motion of the mirrors, which necessitates inertia-free scanning solutions. 
A video frame rate(30 Hz) with 600 $\times$ 300 pixel images requires more than 5~megapixel/second measurement rates. 
Such large frame rates cannot be attained by increasing measurement speed due to limitations imposed by mechanical scanning, as well as pixel dwell time, i.e. signal to noise ratio.
\begin{figure*}[!htbp]
	\includegraphics[width=\linewidth]{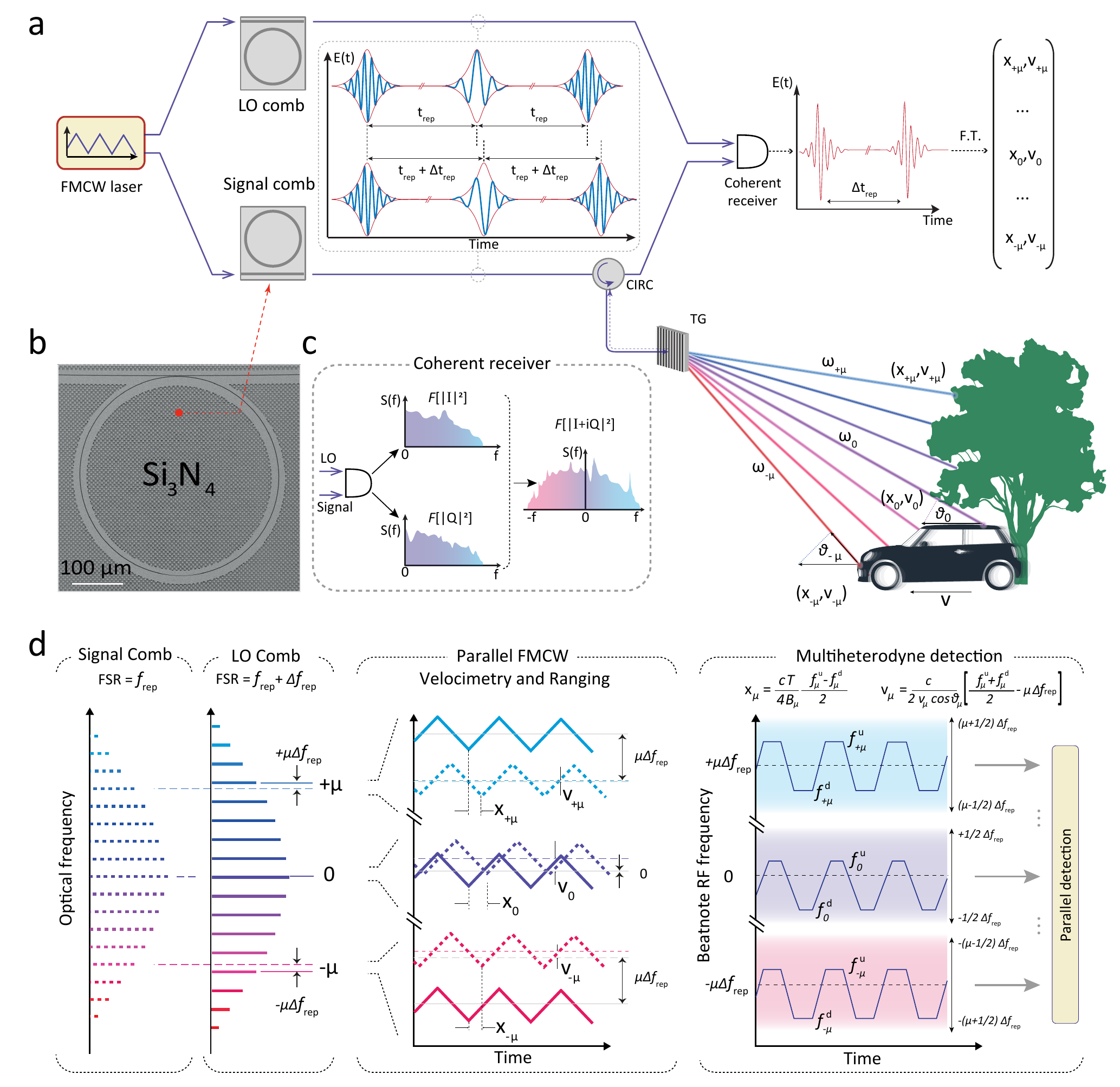}
	\caption{\textbf{Multiheterodyne parallelization of coherent laser ranging.}
	a)~Architecture of the multiheterodyne parallel FMCW LiDAR. A single pump laser with triangular frequency modulation is split, and drives two distinct optical microresonators with slightly different radii, which serve as signal and local oscillator (LO) in the experiment. 
	The signal comb is spatially dispersed over the target area using diffractive optics. Each signal comb tooth $\mu$ represents an independent FMCW ranging channel measuring distance $x_{\mu}$ and velocity  $v_{\mu}$. 
	All channels are simultaneously superimposed with the LO comb on a coherent receiver.
	The interferogram is processed via short-time Fourier transform analysis to retrieve distances $x_{\mu}$ and velocities  $v_{\mu}$.
	b)~Electron microscope picture of 228.43~$\mu$m Si$_3$N$_4$ microring resonator.
	c)~The complex RF spectrum is retrieved by phase diversity detection and Fourier transform.
	d)~Principle of multiheterodyne ranging and velocimetry. The Signal and LO combs have repetition rates $f_\mathrm{rep}$ of 98.90~GHz and 99.39~GHz, respectively. The reflected signal comb light is both time delayed, and frequency shifted due to the Doppler effect. 
	Beat notes of consecutive comb tooth pairs are spaced 490~MHz in the RF spectrum.
	Triangular frequency modulation maps the distance of target objects to two RF beat notes, $f^\mathrm{d}_\mu$ and $f^\mathrm{u}_\mu$, spaced around the center frequency of the multiheterodyne channel $\mu\cdot f_\mathrm{rep}$ offset by the Doppler shift caused by the relative velocity of LiDAR transmitter and target.
	}
	\label{fig_concept}
\end{figure*}
A manifold of solutions for inertia-free scanning based on photonic switching networks \cite{Martin2018}, focal plane arrays \cite{Rogers2020}, spectrally encoded spatial scanning with broadband \cite{Bao2019, Wang2018, Jiang2020} or frequency swept light sources \cite{Qian2020, Okano2020}, or optical phased arrays \cite{Poulton2020} 
have been implemented. 
Yet to date, megapixel rate coherent LiDAR has not been demonstrated.
To further push the acquisition rates, the parallelization technique can be utilized. It has successfully been employed in time-of-flight LiDAR supporting the operation of up to 128 channels.
Recently the parallelization of FMCW has been shown to be possible using dissipative Kerr solitons (DKS) generated in photonic integrated microresonators~\cite{Kippenberg2018} (although electro-optical combs are equally suitable \cite{Zhang2019,Kuse2019}).
DKS, a parametrically driven coherent frequency comb, has an ability 
to faithfully transfer the time-frequency characteristics of 
an FMCW pump laser to all comb teeth at modulation speeds up to 10 MHz with mode spacing of 100 GHz that can be readily multiplexed \cite{Riemensberger2020}. 
The large comb spacing facilitates the spatial separation of the comb teeth with diffractive optics and each tooth can \textit{independently} and \textit{simultaneously} measure the distance and velocity of a target in a truly parallel fashion. 

To unlock the potential of massively parallel FMCW ranging, requires overcoming the challenge that the number of optical balanced photo-receivers, amplifiers and analog-to-digital converters is identical to the channel number that is  employed, which necessitates custom, large-area silicon photonic solutions \cite{Martin2018,Rogers2020} in addition to multiplexers. 

\subsection{Concept of hardware efficient megapixel coherent ranging}
Here we overcome these limitations and demonstrate a hardware-efficient massively parallel coherent FMCW LiDAR based on multiheterodyne mixing of two photonic chip-based soliton microcombs on a single coherent photoreceiver (cf.~Fig.~\ref{fig_concept}), enabling \emph{bona fide} 5.6 megapixel/second measurement rate, with more than 64 simultaneous channels. 
Our approach is a swept frequency version of multiheterodyne detection \cite{Schiller2002} of optical frequency combs, commonly referred to as dual-comb spectroscopy. This technique has attained widespread attention and application in (nonlinear) optical and THz spectroscopy \cite{Keilmann2004,Coddington2016, Hsieh2014}, optical microscopy \cite{Mizuno2021, Hase2018}, distance measurement~\cite{Coddington2009,Trocha2018,Suh2018a}, two-way time-frequency transfer \cite{Giorgetta2013} and microwave photonics \cite{Kim2016}, multi-dimensional spectroscopy \cite{Lomsadze2018} and coherent anti-Stokes Raman imaging \cite{Ideguchi2013a}. By radio-frequency multiplexing, this method enables to decode all individual low frequency (MHz bandwidth) channels using a single high speed (GHz) coherent 'intradyne' detector.
Integrated high-bandwidth coherent receivers are nowadays widespread in data centres around the globe \cite{Doerr2015}. A recently introduced silicon photonics based coherent optical pluggable transceiver 400ZR supports 64~GBaud/s modulation speeds \cite{Isono2021}, which would constitute an off-the-shelf component solution for a chip-scale FMCW LiDAR.
Frequency swept dual-comb with electro-optic modulators has been recently successfully demonstrated in high resolution, spectrally interleaved broadband spectroscopy \cite{Nishikawa2019, Xu2021}.

In our experiments, we utilize a single highly coherent FMCW laser that is amplified, split and coupled into two size-mismatched photonic chip based integrated Si$_{3}$N$_{4}$ \cite{Moss2013} microring resonators (cf.~Fig.~\ref{fig_concept}b) driving two dissipative Kerr solitons \cite{Herr2014}. 
As recently shown, fast frequency tuning of the pump laser within the soliton existence range, retains the solitonic state \cite{Riemensberger2020} in both resonators. 
The rapid frequency modulation is encoded onto the carrier-envelope frequency $f_{\mathrm{ceo}}$ of the pulse, while the pulse repetition rate $f_{\mathrm{rep}}$ remains almost constant. 
The schematic of our architecture is depicted in Fig.~\ref{fig_concept}a. 
The amplified signal comb is dispersed using a transmission grating (966 lines/mm).
In the frequency domain (cf.~Fig.~\ref{fig_concept}d), we obtain two soliton microcombs with slightly different comb line spacing $\Delta f_{\mathrm{rep}}$ where each comb line inherits the pump laser frequency modulation. 
Multiheterodyne mixing of the reflected signal comb with the local oscillator (LO) comb (cf.~Fig.~\ref{fig_concept}c), using a \emph{single} coherent photoreceiver records the complex RF spectrum (cf.~Fig.~\ref{fig_concept}c,d), which allows the distance $x_\mu$ and velocity $v_\mu$ to be recorded for \emph{each comb line $\mu$ simultaneously} ($\mu$ denotes the relative mode number with respect to the pump laser mode).
In comparison to conventional FMCW LiDAR \cite{Pierrottet2008}, multiheterodyne detection modifies the formulas to calculate $(x_\mu, v_\mu)$ from the beat notes $f^\mathrm{u}_{\mu},f^\mathrm{d}_{\mu}$ measured during the up- and down-chirping of the FMCW laser, because the intermediate frequency is no longer at baseband.
Instead, consecutive channels in the radio-frequency (RF) domain are separated by the difference in comb line spacing $\Delta f_{\mathrm{rep}}$. 
To mitigate the degeneracy in optical detection between $+\mu$ and $-\mu$ comb lines located symmetric about the pump ($\mu=0$), we employ a phase diversity receiver architecture \cite{Gao2012,Derr1991} (widely used in coherent telecommunication) and measure both the in-phase (I) and quadrature (Q) components of the multiheterodyne beat note (cf.~Fig.~\ref{fig_concept}c).
The distinction between positive and negative frequencies in the multiheterodyne beat spectrum is obtained via Fourier transform of the complex field amplitude I+iQ \cite{Kikuchi2015}.
The Doppler shift is observed by a deviation of the beat note pattern from $\mu\cdot\Delta f_{\mathrm{rep}}$ and non-zero detection distance translates into a splitting of the RF beat note
\begin{align}
	x_\mu &= \dfrac{cT}{4B_\mu} \cdot \dfrac{f^\mathrm{u}_{\mu} - f^\mathrm{d}_{\mu}}{2} \notag \\
	v_\mu &= \dfrac{c}{2 \nu_\mu cos\theta_\mu} \cdot \left[ \dfrac{f^\mathrm{u}_{\mu} + f^\mathrm{d}_{\mu}}{2} - \mu\Delta f_\mathrm{rep}\right] ,
\end{align}
where $v_\mu cos\theta_\mu$ is a projection of the target velocity on a comb line and $\nu_\mu$ is an optical frequency of the $\mu$-th comb line.


\begin{figure*}[!htbp]
	\includegraphics[width=\linewidth]{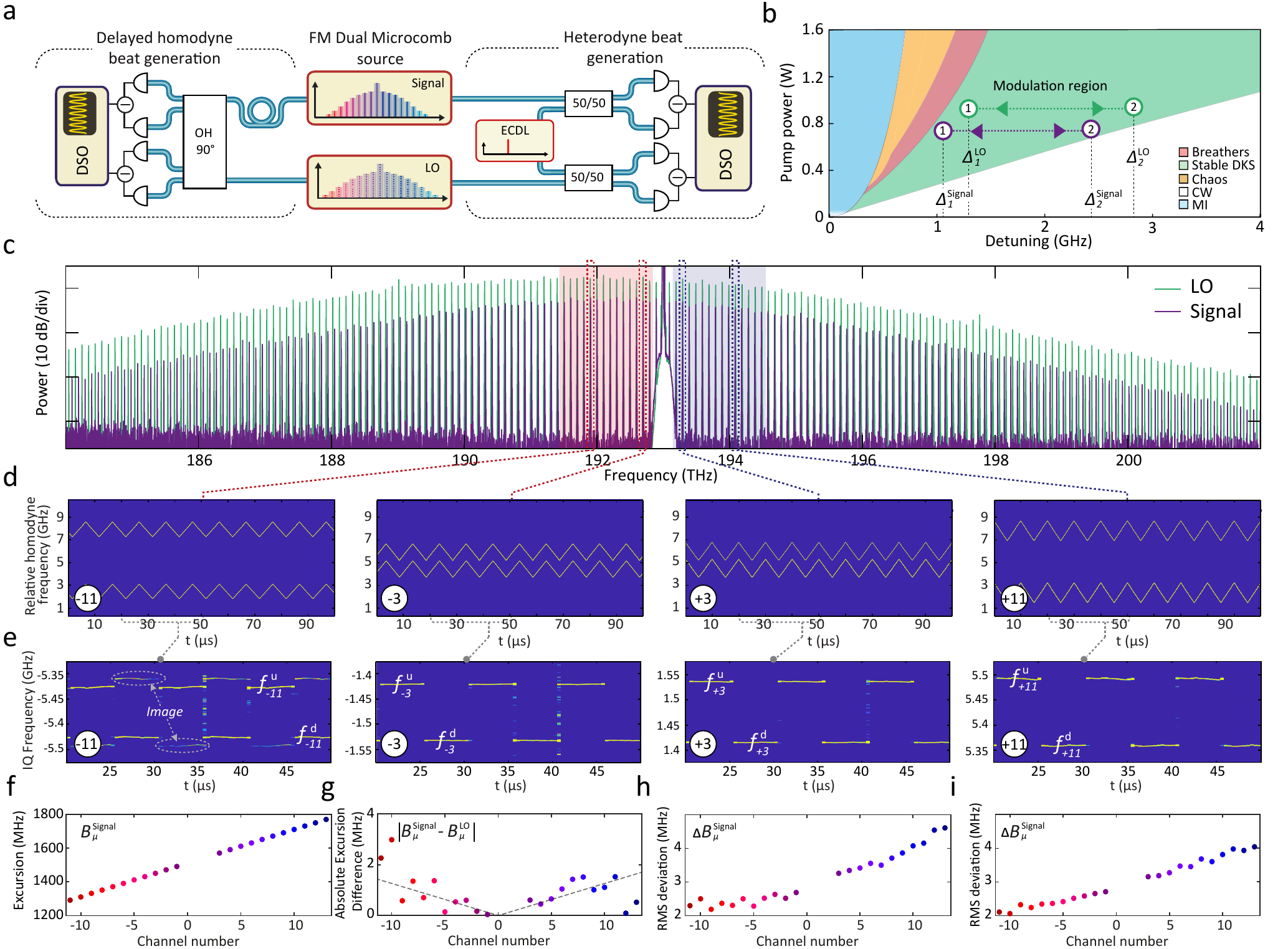}
	\caption{\textbf{Coherent detection of multiple FMCW laser channels.}
	a)~Optical setup for heterodyne and delayed homodyne beat note measurement. The signal microcomb is delayed and mixed in a 90$^\circ$ optical hybrid and superimposed with the local oscillator (LO) microcomb on a pair of balanced photoreceivers for delayed homdyne beat frequency generation. Alternatively, the tuning of individual comb line pairs are characterized by heterodyne mixing with an external cavity diode laser (ECDL).
	b)~Soliton microcomb stability chart. Optical resonances are thermally tuned to superimpose the laser-cavity detuning regions of stable soliton generation (green shaded). MI: modulation instability; CW continuous-wave; 		
	c)~Optical spectra of signal and LO combs featuring a slight offset in repetition rate. 
	The blue(red) shading highlights positive(negative) channels ($\mu=0$ denotes the pump laser) used in LiDAR experiments limited by the available amplifier bandwidth and amplified spontaneous emission noise. 
	d)~Time-frequency map of heterodyne beat spectroscopy at the $\pm3$ and $\pm11$ channels. ENBW 2.45~MHz.
	e)~Time-frequency map of delayed homodyne beat spectroscopy. ENBW 2.45~MHz. 
	f)~Channel-dependent frequency excursion bandwidth at 100-kHz modulation frequency.
	g) Channel-dependent absolute excursion difference of signal and LO combs.   
	h,i) Channel-dependent RMS deviation from a perfect triangular frequency chirp for LO and signal. 
	}
	\label{fig_iq}
\end{figure*}

Figure \ref{fig_iq}a depicts optical spectra of LO and signal combs which feature 99.39 GHz and 98.9 GHz repetition rates, i.e. a 490 MHz difference.
We generate simultaneously two DKS from a single triangularly chirped laser, with an amplitude $B = 1.5$~GHz and period $T = 10 \,\mu$s. This is achieved by thermally tuning the two pump resonances into degeneracy, such that their trajectories in the soliton existance range are similar (cf.~Fig \ref{fig_iq}~b). 
The red and blue shaded spectral regions indicate the comb lines that are evaluated in the ranging experiment. 
We filter out the pump and $\pm1$ sidebands of the LO to remove excess amplified spontaneous emission noise around the laser pump. The pump laser frequency is modulated using a dual Mach-Zehnder modulator biased to single sideband modulation, which in turn is driven by a voltage-controlled digitally predistorted linearized triangular waveform.
First, a heterodyne beat note is obtained by superimposing the frequency combs individually with an external-cavity diode laser onto two balanced photoreceivers (BPD). 
The resulting signals are added and analyzed by short-time Fourier transform. Fig. 2d shows the laser frequency of two individual comb teeth of the two simultaneously chirped local oscillator and signal microcomb for channels $\mu=\pm3,\pm11$, highlighting the similarity and relative spacing of the frequency modulation pattern. 
The delayed homodyne beat note spectrum for the same channels is depicted in Fig.~\ref{fig_iq}e. 
Image peaks are related to imperfect phase compensation in IQ-detection, which is outlined in detail in the methods section.
From a technical point of view, one difference to our earlier demonstration of single comb parallel FMCW LiDAR \cite{Riemensberger2020} is that the signal and LO splitting takes place before soliton generation and the nonlinear microresonators are part of the measurement Mach-Zehnder interferometer (MZI). This enables to employ dual comb heterodyne detection on a single photoreceiver, which is not possible using a single comb only.
The maximum number of LiDAR channels is limited by the optical amplification and coherent receiver bandwidths $B_\mathrm{pd}$. 
The latter limitation reads as $\mu \Delta f_\mathrm{rep} < B_\mathrm{pd}$ and can be overcome by reducing the repetition rate difference $\Delta f_{\mathrm{rep}}$ with the trade-off of a reduced distance ambiguity range.
The channel-dependent frequency excursion $B_\mu$ (cf.~Fig.~\ref{fig_iq}f) is related to the soliton self-frequency shift induced by intrapulse Raman scattering \cite{Karpov2016,Yi2016} and dispersive wave recoil \cite{Brasch2016,Yi2017} and ranges from 1.3~GHz to 1.8~GHz which corresponds to a native distance resolution $\Delta x_\mu  = c/2B_{\mu}$ of 12~cm to 8~cm.
The performance of dual-comb FMCW heterodyne detection relies not only on the mutual coherence of the Signal and LO combs (ensured by degenerate pumping scheme), but also on the equality (cf.~Fig.~\ref{fig_iq}g) and low non-linearity (cf.~Fig.~\ref{fig_iq}h,i) of the chirp transduction from the pump laser. The relative phase deviation between the corresponding Signal and LO comb lines affects the resulting RF signal beatnote linewidth broadening and thus the LiDAR performance (outlined in the methods section).
 


\begin{figure*}[!htbp] 
	\includegraphics[width=\linewidth]{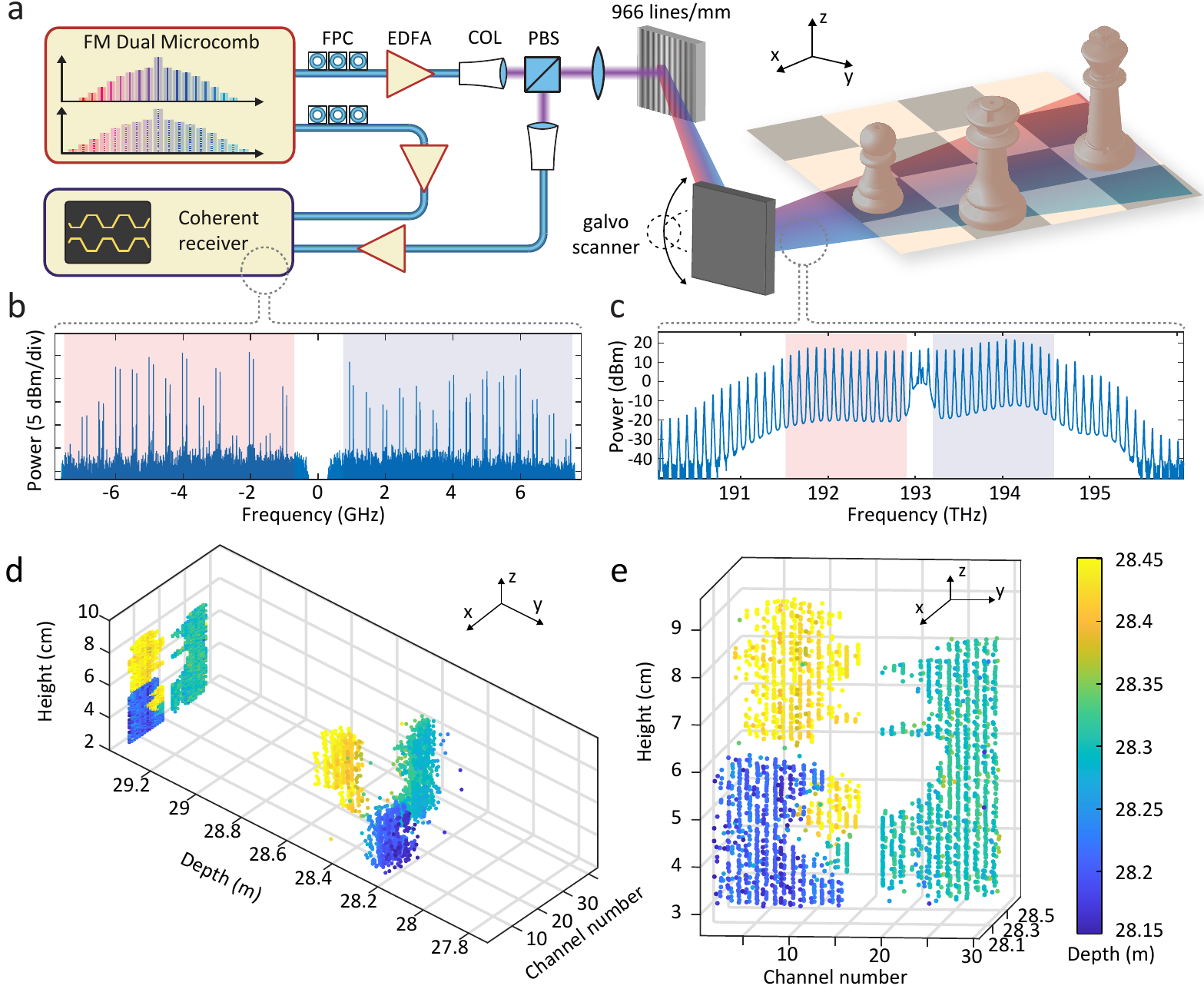}
	\caption{\textbf{Dual comb parallel 3D imaging.} 
	a)~Experimental setup. The amplified signal comb is dispersed in free-space by a 966 lines/mm transmission grating and vertical scanning is provided by a mirror galvanometer. FPC: Fiber polarization controller; COL: collimator; PBS: Polarizing beamsplitter. 
	b)~Power spectral density of the electrical signal obtained at the coherent receiver highlighting 28 FMCW channels. The red and blue shading highlights signals obtained from negative and positive channel numbers, respectively. The resolution bandwidth equals to 100~kHz. SNR ranges in between 5-20~dB with polarization dependent variations.
	c)~Optical spectrum used for 3D imaging. The red and blue shaded regions correspond to the signal plotted in panel b). 
	d,e) Point clouds of the three chess figures obtained during a scan (28 $\times$ 136 points) of the mirror galvanometer. 
	}
	\label{fig_ranging}
\end{figure*}

\subsection{Massively parallel coherent ranging}
To demonstrate the capabilities of dual-comb massively parallel coherent imaging, we perform proof-of-principle parallel ranging experiments. 
The setup is depicted in Fig. \ref{fig_ranging}a. The soliton microcombs are amplified in erbium-doped fiber amplifiers (EDFA). 
The target is composed of three chess figures (queen, king and pawn) placed approximately $\sim$1 m in front of the beam-splitter and optical transmission grating. 
A single-axis galvanometric mirror is used for beam scanning in the vertical direction. Fig. \ref{fig_ranging}c depicts the optical spectrum of the signal comb interrogating the target. 
Fig. \ref{fig_ranging}b depicts the complex RF spectrum obtained by Fourier transform of the I+iQ signal from the coherent receiver of one period acquisition time (10~$\mu$s). Blue and red shadings highlight positive $\mu>0$ and negative $\mu<0$ frequency comb teeth with respect to the pump laser frequency. 
The details of the multichannel data segmentation filtering, IQ phase and amplitude imbalance compensation, and  computational complexity of the required signal processing  are outlined in the methods section. 
Two different projections of the 3D-imaging results for a scan of 136 vertical angles across the set of chess figures are depicted in Fig. \ref{fig_ranging}d,e. A line of 28 pixels is recorded during a single 10~$\mu$s triangular laser chirp, which equates to a \emph{true} , i.e. \emph{bona fide} 2.8 MPix/s coherent distance sampling rate at the sampling oscilloscope. We emphasize that this operation is to well be distinguished from experiments, which do not detect, digitize and process each individual channel that is de-multiplexed, and are thus reporting aggregated data or sampling rates only \cite{Riemensberger2020}. The details on precision, accuracy and repeatability of distance measurements are given in the methods section.

\vspace{-6.0mm}
\subsection{Dense megapixel-per-second coherent ranging}
\begin{figure*}[!htbp] 
	\includegraphics[width=\linewidth]{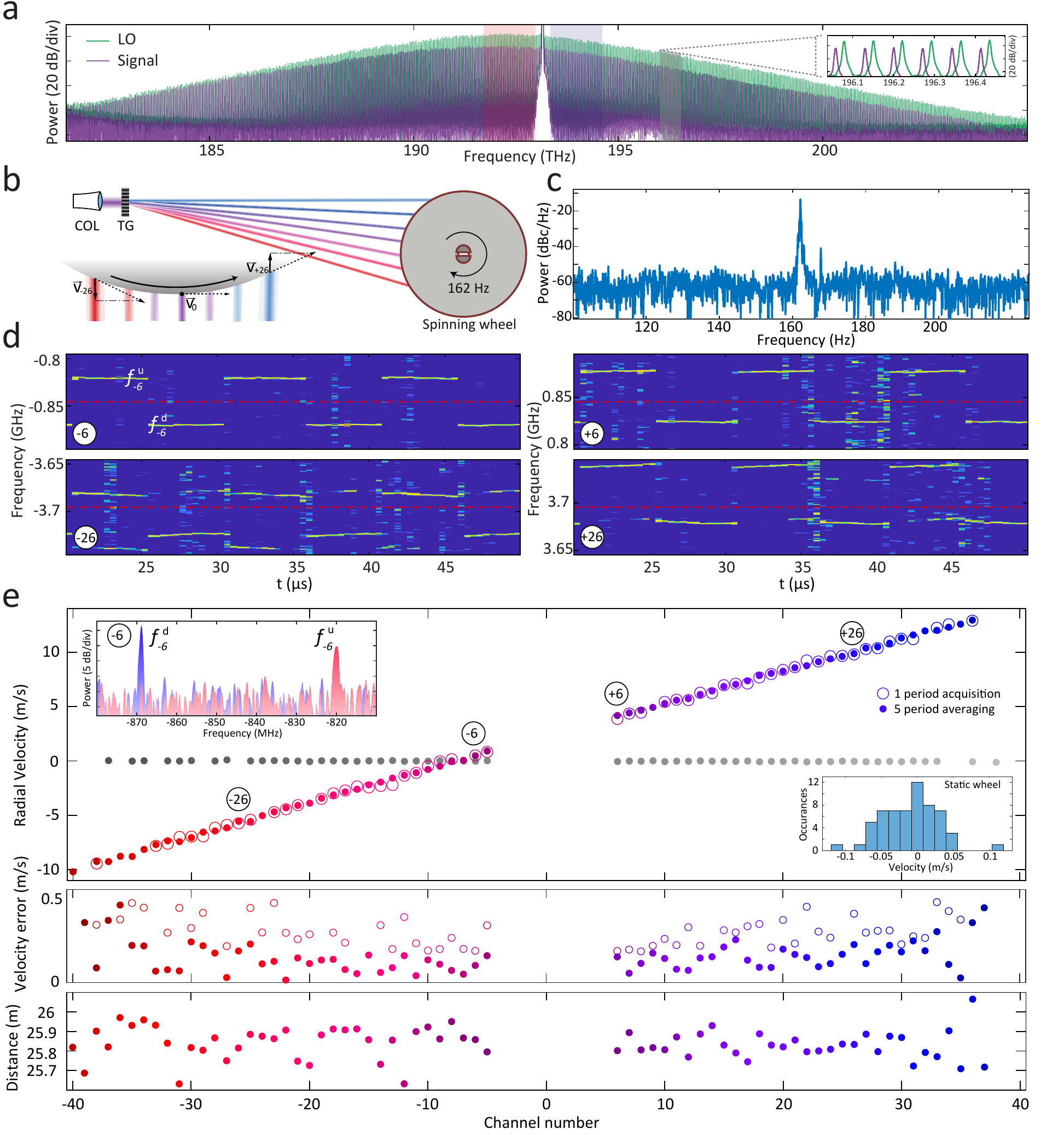}
	\caption{\textbf{Dense dual comb parallel velocimetry measurement at 6.4 megapixel/second rates.}
	a)~Optical spectra of the 35~GHz dual FMCW combs. 
	b)~Schematic illustration of the flywheel section irradiated by the signal comb lines. (Left) COL: Collimator; TG: transmission grating. (Right) The projections of the velocity $v_{\mu}$ of the wheel onto the comb lines.
	c)~Periodogram of the spinning flywheel sound recorded on a cellphone microphone depicting a peak corresponding to the rotation frequency.
	d)~Time-frequency map featuring an offset of the mean beat frequency from the $\mu \Delta f_\mathrm{rep}$ due to the Doppler shift. ENBW 2.45~MHz.
	e)~(Top) Multichannel velocity measurement for the flywheel rotating at around 200~Hz for a single 10~$\mu$s scan (open circles) and five frame stacking (filled circles). Grey data points show velocimetry results for the static wheel. (Middle) Error of velocimetry for single scan and 5 frame stack. (Bottom) Ranging results for five frame stacking.
	}
	\label{fig_velo}
\end{figure*}

In a second proof-of-principle experiment, we demonstrate dense coherent hardware-efficient parallel velocimetry with our dual frequency-modulated soliton microcomb platform at even higher rates of 6.4~megapixel per second.
Lowering the repetition rate to 35 GHz only, illustrates the advantage of the dual comb approach, which alleviates the need to operate at large line spacings compatible with multiplexers and facilitates mode spacing related limitations of channel isolation. 
To this end, we employ two low repetition rate solitons microcomb operating at $f_\mathrm{rep}$ of 35 GHz 
allowing to have more than 60 channels within our EDFA gain window.
While spectral compression of the microcomb from 99~GHz to 35~GHz does limit the triangular chirp frequency excursion to 700 MHz (while preserving the same pump power), this does not limit the accuracy of Doppler velocimetry and we can retain the 100~kHz modulation frequency.
The optical spectra of the signal and LO solitons are shown in Fig. \ref{fig_velo}a. The inset highlights the repetition rate offset $\Delta f_\mathrm{rep}$ of only $\sim$ 140~MHz.
The signal comb is dispersed by the same transmission grating along the circumference of a 20~mm flywheel rotating at 162~Hz (cf.~Fig.~\ref{fig_velo}b,c). 
The pump channel is approximately aligned at the center of the flywheel such that negative channels record an approaching target and positive channels a receding target. 
The time frequency maps of the complex spectrum for the $\mu=\pm6,\pm26$ channels are plotted in Fig.~\ref{fig_velo}d and the dashed red lines highlight the baseband frequencies of multi-heterodyne detection equal to $\mu \Delta f_\mathrm{rep}$.
We calculate the velocities by computing the mean deviation of the beat notes from the equivalent baseband 
and depict the results in Fig.~\ref{fig_velo}e. 
Open circles depict the results after analyzing a single scan period, while the filled circles depict the results obtained over 5 frames averaging.
On average, we attained 56 pixel detections over one period resulting in 5.6 MPix/s, i.e. actually detected, velocity and distance information acquisition speed.
The grey circles depict the velocimetry result for the static wheel, while the measurement uncertainty is depicted in Fig.~\ref{fig_velo}e middle panel and is limited by the mechanical vibrations of the flywheel.
The distance measurement is depicted in Fig.~\ref{fig_velo}e bottom panel. With recent demonstrations of DKS in low-repetition rate microresonators \cite{suh2018gigahertz,Liu2020b} the approach could readily exceed 10 MPix/s.

\subsection{Summary}
In summary, we have demonstrated a megapixel-rate parallel coherent laser ranging based on multiheterodyne detection of chirped carriers on a single coherent receiver.
Two integrated soliton microcombs driven by the same chirped pump laser provide a minimalist implementation of the dual chirped-comb system. The approach is free of channel separation, photo-detection and processing of individual channels. 
Utilization of arbitrary, and in particular very dense, frequency comb channel spacing is possible since multiplexers are not required.
When combined with phase arrays, or other compact non-inertial scanning solutions, our approach provides a route to field deployable MPix/s LiDAR systems, that provide sufficient frame rate to allow video rate 3D imaging. Moreover, high-bandwidth I-Q detectors are already offered commercially in silicon photonics - making our method fully compatible with photonic integration.
A recently demonstrated full heterogeneous integration combining InP/Si semiconductor lasers and ultralow-loss silicon nitride microresonators for DKS generation~\cite{Xiang2021} is feasible as a path to chip-scale parallel FMCW LiDAR. Such photonic integration does not only bring another degree of miniaturization and possibility of wafer-scale production, but also reduces optical loss, increases noise performance of the laser and achievable scanning rates \cite{Lihachev2021}.
Erbium-doped fiber amplifiers could be replaced by broadband semiconductor optical amplifiers (SOA) co-integrated on the silicon substrate \cite{de2020heterogeneous,Vallaitis2010}. It should be noted that SOAs are subject to high nonlinearities and spectral distortion, which however can be reduced by selecting gain media with low $\alpha$-factor \cite{Juodawlkis2009,Huang2011}. 
The broadband amplification would allow to increase comb repetition rate while maintaining high channel count, which leads to improved soliton comb line power and relaxes the requirements on the grating line density. 
Synchronous tuning of the pump laser and the microresonator, i.e. using monolithically integrated piezoelectrical frequency tuners \cite{Liu2020} or frequency comb generation in electro-optical materials \cite{Zhang2019}, serve to completely eliminate the residual nonlinearities of tuning that arise from the Raman self-frequency shift of the soliton and remove the requirement of high-power pumping while possibly extending the soliton existence range.
Finally, and equally important, we believe that the first demonstration of frequency modulated dual solitons might be extended to other  frequencies and will be applicable in the neighbouring fields of spectroscopy and optical coherence tomography (OCT).

\section{Acknowledgements}

This material is based upon work supported by the Air Force Office of Scientific Research (AFOSR), Air Force Material Command, USAF under Award No. FA9550-15-1-0250. J.R. acknowledges support from the EUs H2020 research and innovation program under Marie Sklodowska-Curie IF grant agreement No. 846737 (CoSiLiS). The authors thank Jeremie Renaudier of Nokia Bell labs for valuable suggestions. The Si$_3$N$_4$ samples were fabricated in the EPFL Center of MicroNanoTechnology (CMi). The work was moreover supported by the Swiss National Science Foundation (SNF).

\section{Data Availability Statement}
All data, figures and analysis code will be published on \texttt{Zenodo} upon publication of the work.

\section{Author contributions}
A.L. and J.R. conducted the various experiments and analyzed the data. A.L. designed the samples. J.L. fabricated the samples. All authors discussed the manuscript. A.L.,J.R.,M.K. and T.J.K. wrote the manuscript. T.J.K. supervised the work.

\bibliographystyle{apsrev4-1}
\bibliography{citations}

\section{Methods}

\subsection{Sample fabrication}
\noindent
Integrated Si$_{3}$N$_{4}$ microresonators are fabricated with the photonic damascene process~\cite{Pfeiffer2018d}. Features are defined using deep-ultraviolet (DUV) stepper lithography~\cite{Liu2018} and reactive ion etching and silica preform reflow~\cite{Pfeiffer2018} prior to deposition reduces scattering losses. 
The waveguide width is 1.5~$\mu$m and its height is 0.82~$\mu$m, which leads to a anomalous second order dispersion of $D_{2}/2\pi = 1.13$~MHz and the  third-order dispersion parameter is $D_{3}/2\pi = 576$~Hz. 
The positions of the resonance frequencies close to the pumped resonance are expressed with the series $\omega_{\mu} = \omega_0 + \sum_{i \geq 1} D_i \mu_i / i!$. 
The ring radius of the signal comb is 228.43~$\mu$m and results in a resonator free-spectral-range of $D_{1}/2\pi = 98.9$~GHz. 
The LO comb has similar cross-section and its radius is 227.27~$\mu$m, which leads to a free-spectral-range of 99.35~GHz. 
Both resonators are operated in the strongly overcoupled regime with intrinsic loss rate $\kappa_0/2\pi = 15$~MHz and bus waveguide coupling rate $\kappa_\mathrm{ex}/2\pi = 130$~MHz in order to optimize comb output power and optical signal-to-noise-ratio after amplification. 
The radii of the 35~GHz samples are 645~$\mu$m and 648~$\mu$m resulting in $\Delta f_\mathrm{rep} \approx$ 140~MHz.

\subsection{Dual FM soliton microcomb generation}
\noindent
We use an external cavity diode laser (ECDL) for our proof-of-principle, which is coupled to a dual Mach-Zehnder Modulator driven by a frequency-agile VCO and biased for single-sideband modulation. The laser is amplified, split and coupled into two photonic chips using lensed fibers and double inverse tapers \cite{Liu2018}. Optical power incident on the each chip is $\sim$2~W. Manual tuning with the ECDL piezo is used to tune into resonance. Upon traversing from blue to red, 
 the chaotic modulation instability state collapses into a stable dissipative Kerr soliton  \cite{Herr2014} state. The thermal nonlinearity of Si$_{3}$N$_{4}$ facilities the elimination of undesired multi-soliton states \cite{Guo2017} when  the laser-cavity detuning $\delta^{\mathrm{sig,LO}}$ is reduced beyond the soliton existence range into the transient chaos region \cite{Karpov2019}. If both cavity resonances are aligned, the associated sudden drop in intracavity power increases the detuning $\delta$ of the switching comb, hence leapfrogging the detuning of the non-switching comb. The laser detunings are monitored using the phase modulation response technique introduced in \cite{Guo2017} and temperatures of the samples are adjusted during the dual soliton switching process, if necessary using a thermal tuner. In this way, dual single soliton states can be obtained routinely and quickly. We subsequently optimize the laser-cavity detunings $\delta^\mathrm{sig,LO}$ of signal and LO combs by thermal tuning to minimize the differential Raman shift and hence optimize linearity.

\subsection{Linearization}
\noindent
The precision of coherent laser ranging is directly impacted by the linearity of the chirps that constitute the triangular frequency modulation. The chirp is applied to narrowband CW laser via single side-band modulator driven by a VCO. An iterative linearization algorithm was applied to the pump frequency chirp based on chirp measurement in an imbalanced Mach-Zehnder interferometer (MZI) \cite{Ahn2007}. The detailed procedure can be found in the Methods section of Ref. \cite{Riemensberger2020}.  
We estimate the root mean square nonlinearity (the deviation of the instantaneous frequency from the perfect triangular trace) of the FMCW comb sidebands to be below 5~MHz for sweeping rate of 100~kHz \cite{Riemensberger2020}. The 3dB modulation frequency cut-off that frequency combs are able to reproduce appears to be at least 40~MHz. The maximum tuning rate, i.e. the product of excursion and tuning frequency is determined by the effective filter cavity response time (photon decay rate) and should follow $d \Delta/dt < (\kappa/2\pi)^2 $, where $\kappa$ is a loaded cavity linewidth and equals $\approx$ 150~MHz in our case. 


\subsection{Parallel velocimetry and ranging} 
\noindent
The experimental setup is illustrated in extended data Fig.~\ref{fig_SI_calibration}a. 
The frequency modulation $1/T$ and excursion $B$ of the pump laser are adjusted to 100~kHz and 1.55~GHz, respectively.
A pump laser with triangular frequency modulation drives two distinct optical microresonators with slightly different radii, which serve as signal and LO in the experiment. After filtering (FBG) and amplification (EDFA), the signal comb is spatially dispersed over the target area using diffractive optics (transmission grating~(966~lines/mm)). 
Each signal comb tooth $\mu$ represents an independent FMCW ranging channel measuring distance $x_{\mu}$ and velocity  $v_{\mu}$. 
The reflected signal is post-amplified similar to signals in high-bandwidth and long haul optical telecommunication systems and simultaneously superimposed with the amplified LO comb on a coherent receiver.
A programmable filter (WS) is used to filter out excess (amplified spontaneous emission) ASE noise around the pump of the LO.
A bistatic detection with separate collimators for the transmit and receive path is chosen to minimize spurious backreflection in the fiber components.
Another balanced photodiode is used to calibrate $\Delta f_\mathrm{rep}$ and the channel-dependent frequency excursion $B$.
A mirror galvanometer (shown in Fig. \ref{fig_ranging}) is used to scan the dispersed light beams in the vertical direction.

The imaging experiment (Fig. \ref{fig_ranging}) is carried out with a signal soliton microcomb repetition rate of 98.9~GHz and repetition rate difference of 490~MHz.
The in-phase (I) and quadrature (Q) signals from the coherent photoreceiver are recorded on a fast oscilloscope and processed offline. 
We utilize microcomb channels between  $\mu=-15,+15$, where $\mu=0$ denotes the pump frequency, while $\mu=0,\pm1$ are filtered out by the WS due to the excess of the ASE, resulting in 28 independent channels. The required coherent receiver bandwidth for this configuration is determined as $\mu \Delta f_\mathrm{rep} \approx$ 7.5~GHz. All the data points are collected from a single trace covering 136 vertical lines. 
The galvo-scanner is set on a linear scan mode, while the oscilloscope is triggered in segmential acquisition mode with 10~$\mu$s segment scan time and 10~ms idle time between segments to allow the scanning Galvanometer enough time to rotate and capture the full scene. 
Overall, the measurement time is less than 1.5 seconds and was limited by the galvanometer response bandwidth, not the pixel sampling rate. The 3D point cloud of $136 \times 28$ pixels was obtained after post-processing. 
The test scene is presented in extended data Fig. \ref{fig_SI_chess}. 
Chess figures are placed approximately one meter after the transmission grating, the differential delay between signal and LO combs in optical fiber is around 27~m. 
The Pawn and Queen figurines are spaced 10.5~cm apart while the Queen and King figurines are spaced 13.5~cm apart.  
The relative distance accuracy is depicted on extended data fig. \ref{fig_SI_uncertainty} by plotting histograms of detected pixels for both positive and negative channels. 
To estimate the overall distance precision of the dual comb LiDAR we perform ranging experiment of a static object. Channel-dependent precision is obtained by measuring the distance corresponding to each individual channel during 49 chirp periods (490~$\mu$s) and calculating the standard deviation (extended data fig. \ref{fig_SI_precision}). Precision of our system ranges from 1 to 5.5~cm. 
The precision value is governed by how precisely one can define the RF frequency beatnote. It depends both on the frequency bin spacing determined by the target interrogation time and nonlinearities in the system that broaden the beatnote. In our system, Raman nonlinearity degrades the comb line precision towards higher $|\mu|$ since its impact is proportional to the relative comb number (described below).

The velocimetry measurements are carried out with 35~GHz microcombs and 140~MHz $\Delta f_{\mathrm{rep}}$ allowing us to increase the number of operational channels $\mu = \pm 5,\pm 40$. Channels $\mu = -4,+4$ are filtered out by the WS due to the excess ASE. 
The frequency excursion $B_{\mu}$ ranges from 500 up to 950 MHz (extended data Fig.~\ref{fig_SI_calibration}d), which results in decreased depth resolution compared to the previous 100~GHz soliton microcomb system. 
The flywheel is rotating at frequency 162~Hz resulting in the 20.4~m/s tangential velocity.
The velocity errors depicted in Fig. \ref{fig_velo}e are attributed to the mechanical vibrations of the flywheel. The velocity error of a single measurement is defined by calculating a variance of the Gaussian fitted to the beatnote. For five consecutive measurements a standard deviation of $N \leq 5$ detected velocity values is calculated for every channel.
Mechanical vibrations of the flywheel not only impact distance and velocity precision, but also limit the total number of possible detections. 
In extended data Fig.~\ref{fig_SI_velocity_statistics}, we present the detection statistics obtained over 190 $\mu$s continuous measurement time, i.e. 19 periods of the triangular waveform. 
Unfilled circles in panel a) (the same as in Fig. \ref{fig_velo}d) correspond to a measurement over one period with a mean number of detections equal to 56. 
Filled circles correspond to the data averaged over five periods. Panel b) depicts a probability for each channel to be detected. 
The roll-off at high $\vert\mu\vert$ originates from the limited optical amplification bandwidth. 
Panel c) represents the probability distribution of the sum of successfully detected pixels during individual scan periods, which is calculated by dividing the number of detections in a period over the full number of channels ($\mu = -40,+40$).

\subsection{Coherent detection and post-processing}
\noindent
The frequency of the complex heterodyne beat note for the channel $\mu$ and a photon time-of-flight $\tau$ follows as the difference between the instantaneous optical frequencies of signal and LO comb teeth $\nu_{\mu}^{\mathrm{sig,LO}}$
\begin{equation}
	\begin{gathered}
		f^\mathrm{IQ}_{\mu}(t) =  \nu_{\mathrm{\mu}}^{\mathrm{LO}} - \nu_{\mathrm{\mu}}^{\mathrm{sig}}\\
		= \delta(t) +  \mu f_{\mathrm{rep}}^{\mathrm{LO}}(t)-\delta(t+\tau) -  \mu f_{\mathrm{rep}}^{\mathrm{sig}}(t+\tau).
	\end{gathered}
	\label{eq:f_IQ}
\end{equation}
The first and third terms are similar to the case of single frequency coherent photoreceiver FMCW LiDAR \cite{Gao2012}. In our case that frequency is offset by the repetition rate difference of the soliton microcombs multiplied by the channel number and adding the Doppler shift due to the relative target velocity $v$, we arrive at the following expressions: 
\begin{align}
	f^\mathrm{u}_\mu &= \mu\cdot\left(f_{\mathrm{rep}}^{sig} - f_{\mathrm{rep}}^{LO} \right) + \dfrac{B}{2T} \cdot \tau + \nu_\mu \cdot \dfrac{v}{c} \notag \\
	f^\mathrm{d}_\mu &= \mu\cdot\left(f_{\mathrm{rep}}^{sig} - f_{\mathrm{rep}}^{LO} \right) - \dfrac{B}{2T} \cdot \tau + \nu_\mu \cdot \dfrac{v}{c}. 
	\label{eq:f_ud}
\end{align}

We point out that the linewidth of beatnote $f^\mathrm{IQ}_{\mu}$ depends on the relative phase deviations between the Signal and LO comb lines (extended data figures \ref{fig_SI_rel_phase},\ref{fig_SI_rel_psd}) that inherent the chirp from single FMCW pump laser. 
This places additional requirements on linearity and uniformity of chirp transduction during DKS generation compared to the single comb case (extended data figures \ref{fig_SI_2combs_phase},\ref{fig_SI_2combs_psd}).
 
According to our notation, positive frequencies in the complex RF spectrum correspond to optical carrier frequencies larger than than the pump, negative frequencies to channels with smaller optical carrier frequencies than the pump. 
The distance and velocity can be inferred from these expressions and are depicted in the inset of Fig.~\ref{fig_concept}d. 

Our coherent receiver consists of 90$^\circ$ optical hybrid coupler, two balanced photodetectors and subsequent RF amplifiers. 
This type of receiver is commonly referred as "phase-diversity homodyne receiver" \cite{Hodgkinson1985,Davis1987} or "intradyne receiver" \cite{Derr1991}. 
It is similar to established receivers used in quadrature-amplitude modulation schemes that are employed in long haul optical communication systems. 
It allows full reconstruction of the amplitude and phase of the RF beat note between the signal and the LO and reveals distinct spectral information both in positive and negative RF frequencies. 
Due to the differential delays of discrete components and the response of the balanced photoreceivers and amplifiers, the in-phase (I) and quadrature (Q) signals are not perfectly orthogonal. 
We perform IQ imbalance correction \cite{Ellingson2003}, which substantially improves orthogonality (extended data Fig. \ref{fig_SI_IQ}). However, for higher channels mismatch still exists and one can observe 'images' on time frequency maps (i.e. Fig \ref{fig_iq}e left panel).

The post-processing relies on short time Fourier transform over the half of the period to retrieve $f^\mathrm{u}_{\mu},f^\mathrm{d}_{\mu}$ RF frequencies. 
In this regard, the 'image' peaks do not pose a problem, since we know that the positive frequencies will give a higher frequency beat note first while for the negative ones it would be the lower one (extended data Fig. \ref{fig_SI_IQ}). 
It is obtained by triggering data acquisition on the oscilloscope by function generator used to create the triangular frequency modulation control signal for the VCO.
Distance information is obtained as a difference between $f^\mathrm{u}_{\mu},f^\mathrm{d}_{\mu}$, while the mean offset of $f^\mathrm{u}_{\mu} + f^\mathrm{d}_{\mu}$ from $\mu \Delta f_\mathrm{rep}$ is proportional to the velocity. 
Calculation of the $\Delta f_\mathrm{rep}$ is outlined below.
Further improvements, especially in long range detection can be achieved using active demodulation analysis \cite{Feneyrou2017}.

To evaluate the real-time digital processing requirements, we give an estimate of computational complexity.
The main computational operation is a discrete Fourier transform (DFT), which can be computed via the Fast Fourier transform (FFT) method. The complexity of FFT can be estimated as $\approx$ 4N*log$_2$(N), where N is a sample size \cite{Frigo1998}. 
Considering a coherent receiver of 10~GHz bandwidth detecting 40 comb lines (channels) with an FM period of 10~$\mu$s, one requires 2 DFTs for the up-ramp and down-ramp of the chirp. Let the sampling speed be 20~GS/s fulfilling the Nyquist requirement. The number of the sampled points for one ramp (half of the period) is 20~GS/s * 5$\mu$s = 10$^5$. 
The required number of operations for FFT is ~4N*log$_2$(N) = 6.64*10$^6$.
Next we use Gaussian fitting for peak detection in every particular frequency band. For nonlinear fitting, we take 20 points covering 20 / 5$\mu$s = 20*200~kHz = 4~MHz band, which is broader than the beatnote linewidth.
These operations require a peak search and threshold detection up to 40 times in intervals of 10$^5$/40 points ($O$(N) - complexity) and to fit a Gaussian for 20 points surrounding every maximum ($O$(N$^3$) - complexity). The complexity of the latter 2 operations is more than an order of magnitude lower than the number of operations needed for FFT.
Thus we estimate the computational requirement to be $\approx$1.33 TFlops rate. 
A commercial FPGA Artix-7 by Xilinx (price $\approx$100\$) could handle this task with 930 GMACs ($\approx$1.86 TFlops).

\subsection{Calibration of channel-dependent frequency excursion}
\noindent
In general, the soliton repetition rate $f_\mathrm{rep}$ depends on the laser cavity detuning, because of intrapulse stimulated Raman scattering \cite{Karpov2016,Yi2016} and the soliton recoil effect associated with dispersive wave emission\cite{Brasch2016,Yi2017}. 
During chirped soliton generation, this induces a change of the soliton repetition rate during each chirp cycle, which is observed in the form of a  channel-dependent frequency excursion $B_{\mu}$ and chirp nonlinearity \cite{Riemensberger2020}. We observe that it does not depend strongly on the pump laser detuning and is independent of the pump power.

We measure the time-dependent chirp on both signal and LO combs and all channels by recording a heterodyne beat note with a second laser simultaneously on a pair of balanced photodetectors  and a fast sampling oscilloscope (cf. Fig.~\ref{fig_iq}a). 
For direct comparison of signal and LO chirps, we add the two beat note signals prior to short-time Fourier transform and depict both beat notes on a single panel. 
The full dataset corresponding to the subset of heterodyne beat signals presented in Fig.~\ref{fig_iq}d is depicted in extended data figure \ref{fig_SI_heterodyne}. 
The amplitude of the triangular ramps decreases from positive to negative channels both for LO and signal. 
We retrieve the frequency excursion by fitting a symmetric triangular ramp to the time-dependent frequency of the heterodyne beat notes. 
The result of this analysis is plotted in Fig.~\ref{fig_iq}f-i.

Heterodyne beat spectroscopy is well suited to characterize the chirp waveforms, but not practical for ranging and velocimetry, as it requires an independent reference laser. Hence, during the experiments presented in figures \ref{fig_ranging} and \ref{fig_velo}, we utilize a reference optical fiber MZI that is derived by tapping a fraction of signal and LO and beating them together on a second balanced photodiode  (cf. extended data Fig. \ref{fig_SI_calibration}a). Inphase detection suffices, because Doppler-shifts are negligible in the reference MZI and all channels $\mu$ observe the same distance $x_{\mu}$ but different frequency excursions $B_{\mu}$, which simplifies interpretation of the signal, which is plotted in extended data figure \ref{fig_SI_calibration}b. 

Due to the channel-dependent excursion $\pm\mu$ channel would give a beat note consisting of four lines (cf extended data Fig. \ref{fig_SI_calibration}b), where two outer lines correspond to positive frequencies with higher excursion and two inner lines correspond to negative frequencies with lower excursion. 
Thus, we perform excursion inference for the 6 channels $\mu = \pm1,\pm3$ (that is allowed by 1.6 GHz bandwidth) and further extrapolate it considering linear dependence (cf. Fig. \ref{fig_iq} f,h,i). 
For the 35 GHz chips we utilize channels $\mu = \pm2,\pm10$ for calibration.
Knowledge of the pump excursion enables us to calculate the distance of MZI, while given distance one can obtain excursion for channel $\mu$ from time-frequency analyzes. 
The same way one can infer channel-dependent excursion from real distance measurement if the distance to the target is the same for all of the channels (e.g. carton block). 
Furthermore, we can also calculate $\Delta f_\mathrm{rep}$ from the same calibration measurement. Since there is no Doppler shift, $\Delta f_\mathrm{rep}$ is equal to the mean of two beat notes divided by $|\mu|$. 
Hence, we are able to conduct the experiments with free-running soliton microcombs. 
Extended data Fig. \ref{fig_SI_calibration}c,d present comparison of the excursion inferred from the calibration measurement and further extrapolated and from a real distance measurement with a carton block. 
In the experiments described in the main part of the manuscript we utilized excursion and $\Delta f_\mathrm{rep}$ for velocity and distance calculations inferred from the calibration measurements.

\subsection{Impact of nonlinear chirp transduction}
\noindent
Multiheterodyne mixing of dual-chirped soliton microcombs necessitates not only soliton generation to preserve chirp linearity from the pump to the comb teeth, but also requires that chirps are transduced equally, so as to avoid a difference in the frequency excursion $B_{\mu}$ between the signal and LO comb teeth that would broaden the ranging beatnote and penalize detection precision and sensitivity. 
We have two contributions effecting soliton repetetion rate given change in pump frequency: the chromatic dispersion of the cavity $D_2$, i.e. frequency dependency of the free-spectral-range $D_1$, and Raman shift $\Omega$~\cite{Karpov2016,Yi2016}. The overall change in repetition rate as a function of the detuning $\delta$ (frequency difference between the pump and 'cold' cavity resonance)
can be written as
\begin{equation}
	f_\mathrm{rep}(\delta) =  f_{\mathrm{rep}}(0) + (\delta + \Omega(\delta)) \frac{D_2}{D_1}.
	\label{eq:frep}
\end{equation}
For dielectric integrated microresonators it is generally found that $\delta \ll \Omega(\delta)$ and hence the first term in the round brackets can be omitted. We also neglect the weak dependence of the free-spectral range and dispersion on the laser-cavity detuning that can be derived as $D_n(\delta) - D_n(0) \approx \delta \cdot D_{n+1}/ D_1 \ll D_n(0)$. The linear dependence of the instantaneous $f_\mathrm{rep}$ on the laser-cavity detuning $\delta$ induces a channel-dependent frequency excursion $B_\mu$ without introducing nonlinearity into the triangular chirp \cite{Riemensberger2020}. Non-linear coupling bestween $f_\mathrm{rep}$ and $f_\mathrm{ceo}$ during the detuning sweep due to the soliton self-frequency shift leads to distortions of the triangular chirp. 
The nonlinear relation between the laser cavity detuning and the Raman soliton shift was derived in Ref.~\cite{Yi2016}:
\begin{equation}
\delta = \sqrt{\frac{15c\beta_2\omega_0}{32nQ}\frac{\Omega}{\tau_{R}}}-\frac{c\beta_2}{2n}\Omega^2,
\label{eq:raman}
\end{equation}
where $\tau_{R}$ is Raman shock time and $\beta_2 = -\frac{n}{c}D_2/D_1^2$ is the chromatic dispersion term.
Extended Data Fig.~\ref{fig_SI_Raman}~a,b depict the time-varying detuning $\delta(t)$ of a triangular chirp sequence of the pump laser and the Raman induced soliton self-frequency shift $\Omega/2\pi$ caused by the variation of the detuning. 
Extended Data Fig.~\ref{fig_SI_Raman}~c depicts the induced variation of $f_\mathrm{rep}$ (blue) and $\Delta f_\mathrm{rep}$ component (red).
We investigate the degradation of chirp linearity and similarity between the signal and LO combs due to the Raman effect as function of their respective and in general dissimilar detunings $\delta^{\mathrm{sig,LO}}$ by inserting the Raman induced periodic change of $f_\mathrm{rep}$ into equations \ref{eq:f_IQ} and \ref{eq:f_ud}. 
We distinguish the cases of vanishing and modest differential detunings $\Delta\delta = \delta^{\mathrm{sig}} - \delta^{\mathrm{LO}} = 10 MHz$ (cf.~Fig.~\ref{fig_SI_Raman}~d,e,f).\\
The small difference in $f_{\mathrm{rep}}$ between signal and LO microcombs results in different Raman shifts, which causes frequency excursions to differ between the signal and LO combs on the scale of 1~MHz, which is comparable to the induced chirp nonlinearities of pump to sideband chirp transduction (extended data Fig. \ref{fig_SI_Raman}~c,e,f). 
According to the equations \ref{eq:f_ud} Raman nonlinearity scales with the relative mode number from the pump. This can be directly seen from extended data figures \ref{fig_SI_rel_phase},\ref{fig_SI_rel_psd},  where relative phase deviation between corresponding Signal and LO comb lines and their phase noise power spectral densities (PSD) $S_{\phi\phi}(f)$ increase for higher $|\mu|$. Values in the top right corners of extended data figure \ref{fig_SI_rel_psd} denote integral of the PSD . A value less than $2/\pi$ heuristically corresponds to the Fourier transform limited linewidth, since $\int_{\frac{\Delta\nu}{2}}^{\infty} S_{\phi\phi}(f) \,df = 2/\pi$ \cite{Xie2017}. Increased linewidth of the detected beatnotes results in the reduced resolution and degraded SNR, as well as impaired distance precision (extended data figure \ref{fig_SI_precision}). Substantially higher phase deviations in channels $+11,+3$ are attributed to the presence of mode crossings in the microresonator spectra \cite{Yi2017}. Similarly, phase deviations in channel $-1$ do originate from unfiltered ASE noise.
To determine the distance inference degradation due to the soliton induced Raman self-frequency shift, we simulate the full signal generation and coherent detection chain at a distance of 30~m with a custom \texttt{MATLAB} script and apply the same data analysis techniques to the artificial (i.e. numerically generated) multiheterodyne LiDAR trace as for the real data.
The resulting time frequency traces are depicted in extended data figure~\ref{fig_SI_Raman}~g,h,i for the pump and the 15$^\mathrm{th}$ low and high frequency comb teeth.
The overall manifestation of the soliton Raman self-frequency shift induced nonlinearity can be seen in the tilt and curvature of the time dependent complex RF beat note frequency. The effects of the  nonlinearities are exacerbated for channels further away from the pump.
The curvature that is imposed by the nonlinear component of the Raman frequency shift leads to a bias of the measured distances that increases with length difference between the target and the calibration MZI, differential detuning $\Delta\delta$ and channel number (cf. extended data figure~\ref{fig_SI_Raman}~j,k,l,m). In the ranging experiments presented here, we calibrate the bias by measuring the static target with the same distance for all the channels and apply this correction for further ranging experiments.

In a real-world deployment of our system, a feed-forward scan scheme \cite{Liu2020} or phase-locking the comb to a tuned resonator \cite{Kuse2020} to avoid detuning-dependent changes of the microcomb repetition rates would be favorable, and are possible to avoid this effect.
Co-integration of both microcomb resonators on a single chip to avoid differential frequency drift is equally straightforward.




\subsection{Signal-to-noise ratio of multiheterodyne FMCW LiDAR}
\noindent
A significant advantage of heterodyne detection is that it can approach the photon shot noise limit of detection and attain single photon sensitivities for low signal powers on a conventional semiconductor photodiode, if sufficient LO power is supplied for amplification. 
In the multiheterodyne case all individual channel LO are impinging on a small number of photodiodes (4 in case of a phase diversity receiver with balanced photodiodes).
Generally, for a heterodyne LiDAR detection $P^{\mathrm{LO}} > P^{\mathrm{sig}}$ 
and the shot noise (mainly contributed by the LO) is a dominating source of noise (i.e., higher than the thermal noise) if sufficient LO power is supplied to the photodiode. 
\begin{equation}
\mathrm{SNR} =  \frac{\langle I^2 \rangle}{\langle \Delta I^2 \rangle} = \frac{P^{\mathrm{sig}}}{\hbar\omega B_{\mathrm{RF}}}
\label{eq:SNR_simple}
\end{equation}
Where $B_{\mathrm{RF}}$ is the resolution bandwidth of the analysis FFT.
For simplicity, we consider quantum efficiency of the photodiode to be 100$\%$, thus the electric current is $I = RP = \frac{e}{\hbar\omega}P$.
If an optical amplifier is used, than for the shot noise limited detection the total SNR would be at least 3dB lower:
\begin{equation}
\mathrm{SNR} = \frac{P^{\mathrm{sig}}}{2\hbar\omega B_{\mathrm{RF}}}
\label{eq:SNR_ampl}
\end{equation}
Below we elaborate a more detailed analyzes for the case of multiple channels detected on a single balanced photodiode.
Consider N distinct signal-LO channels distributed over the photodiode bandwidth with equal powers of the signal channels $P^{\mathrm{sig}}_\mu = P^{\mathrm{sig}}$ and LO channels $P^{\mathrm{LO}}_\mu = P^{\mathrm{LO}}$ for $\mu \in [1,N]$. 
For heterodyne detection the photodiode current of channel $\mu$ is $2R^2P^\mathrm{LO}P^\mathrm{sig}$, while the total noise $\langle \Delta I^2 \rangle$ effecting channel $\mu$ would consist of shot noise, thermal noise, spontaneous-spontaneous beating noise, and signal/LO-spontaneous beating noise \cite{Baney2000}.

The photon shot noise is proportional to the mean current impinging on the photodiodes 
\begin{equation}
\begin{gathered}
\sigma^2_\mathrm{sh} =  2q\langle I \rangle B_{\mathrm{RF}}\\
 = 2qR(NGP^{\mathrm{LO}}+NGP^{\mathrm{sig}}+P_\mathrm{ase}^{\mathrm{LO}}+P_\mathrm{ase}^{\mathrm{sig}})B_{\mathrm{RF}}\\
 \approx  2qR N G P^{\mathrm{LO}}B_{\mathrm{RF}}
\end{gathered}
\label{eq:shot_noise}
\end{equation}
mainly contributed by $N$ LOs. 
Where $P_\mathrm{ase}$ is spontaneous emission noise and it equals $\rho_\mathrm{ase}\Delta \nu_\mathrm{amp}=n_\mathrm{sp}\hbar\omega(G-1)\Delta \nu_\mathrm{amp}$ with the amplification bandwidth $\Delta \nu_\mathrm{amp}$ (4~THz for EDFA) and the spontaneous emission factor $n_\mathrm{sp} = 1$ for the amplifier with complete inversion, $q$ is the electron charge.
Additionally, we consider equal gains for the detector pre-amplifiers of the signal and local oscillator microcombs $G^\mathrm{LO} = G^\mathrm{sig} = G$.

We consider the thermal noise of the photodiodes at room temperature
\begin{equation}
\sigma_\mathrm{th}^2 = 4\frac{kT}{Z}B_{\mathrm{RF}}
\label{eq:th_noise}
\end{equation}
where $kT$ is a thermal energy and the impedance $Z$ of the load is 50 Ohm.
Spontaneous-spontaneous and signal/LO-spontaneous beating noises include cross terms only, because we employ balanced photodetection,
\begin{align}
\sigma_\mathrm{sp-sp}^2 &= 4R^2\rho^{\mathrm{sig}}_\mathrm{ase}\rho^{\mathrm{LO}}_\mathrm{ase} \Delta \nu_\mathrm{amp} B_{\mathrm{RF}}, \\
\sigma_{\mathrm{sig/LO}-\mathrm{sp}}^2 &= 4R^2(NGP^{\mathrm{LO}})\rho^{\mathrm{sig}}_\mathrm{ase} B_{\mathrm{RF}} \\
&+ 4R^2(NGP^{\mathrm{sig}})\rho^{\mathrm{LO}}_\mathrm{ase} B_{\mathrm{RF}}.
\label{eq:spontaneous_noise}
\end{align}
In the experiment, we use two optical amplifiers for the signal comb: a booster-amplifier before free-space emission and a detection pre-amplifier after light collection. 
In the above equations, $P^{\mathrm{sig}}$ denotes the optical power of the signal comb line at the input of the 90$^\circ$ optical hybrid. 
The ASE noise of the booster amplification stage is irrelevant in the coherent ranging application, as the optical loss in free space ($>60$~dB) generally surpasses the single stage amplification gain. 
Thus the photon shot noise would be dominating at the post-amplification stage.
Combining all noise terms and neglecting minor contributions, we determine the SNR for a given channel:
\begin{align}
\mathrm{SNR}_\mu &=  \frac{\langle I_\mu^2 \rangle}{\sigma_\mathrm{sh}^2+\sigma_\mathrm{th}^2+\sigma_\mathrm{sp-sp}^2
+\sigma_{{\mathrm{sig/LO}}-\mathrm{sp}}^2}\\
 &\approx  \frac{2R^2P^{\mathrm{sig}} GP^{\mathrm{LO}}}{4R^2(NGP^{\mathrm{LO}})\rho^{\mathrm{sig}}_\mathrm{ase} B_{\mathrm{RF}}} 
\approx \frac{P^{\mathrm{sig}}}{2N\hbar\omega B_{\mathrm{RF}}}
\label{eq:SNR_final}
\end{align}
Thus, in the case of shot-noise limited operation, the SNR of a channel $\mu$ is reduced by the shot noise of the additional local oscillators and $N$ times lower than in case of its detection on $N$ separate photodetectors. This multiheterodyne penalty is well known in the realm of dual comb spectroscopy \cite{Newbury2010} and amounts to 17~dB for 50 channels. It can be reduced by spectral slicing or interleaving.
While the approach comes at the expense of reduced SNR due to the multiheterodyne detection penalty, the latter is compensated for by the absence of multiplexers or photonic integrated solutions for detection of individual channels, which typically exhibit significant insertion loss. 


\section{EXTENDED DATA FIGURES} 
\renewcommand{\figurename}{\textbf{Extended Data Fig.}}
\setcounter{figure}{0}

\newpage
\begin{figure*}[!htbp] 
	\includegraphics[width=\linewidth]{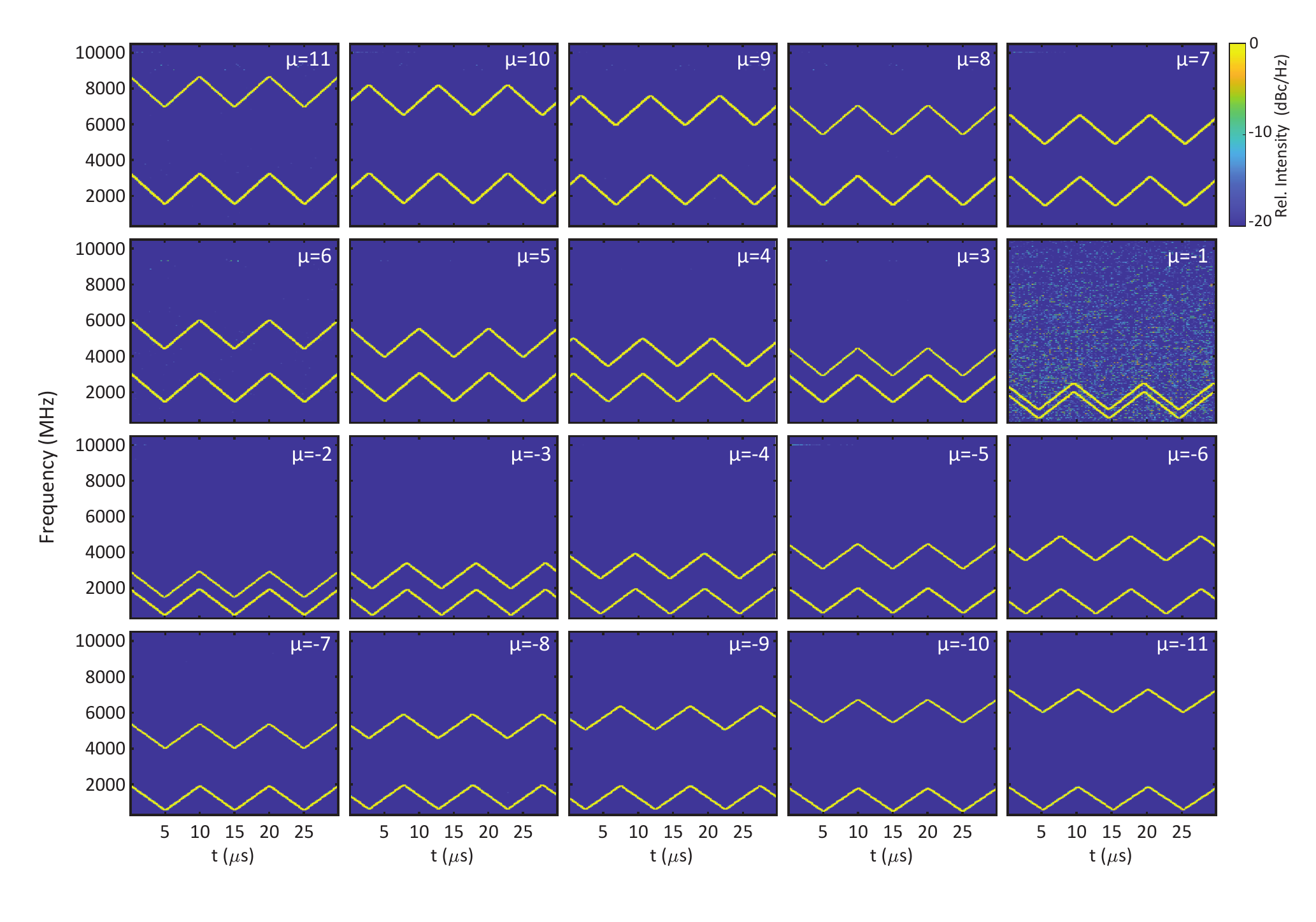}
	\caption{\textbf{Heterodyne measurement of FMCW Dual-Comb.}
	Time-frequency map of heterodyne beat spectroscopy of signal and LO microcombs with external reference obtained by short-time Fourier transform with resolution bandwidth 2.45~MHz. Depicted here is the full data set corresponding to Fig.~\ref{fig_iq}d of the main manuscript.
	}
	\label{fig_SI_heterodyne}
\end{figure*}

\newpage
\begin{figure*}[!htbp] 
	\includegraphics[width=\linewidth]{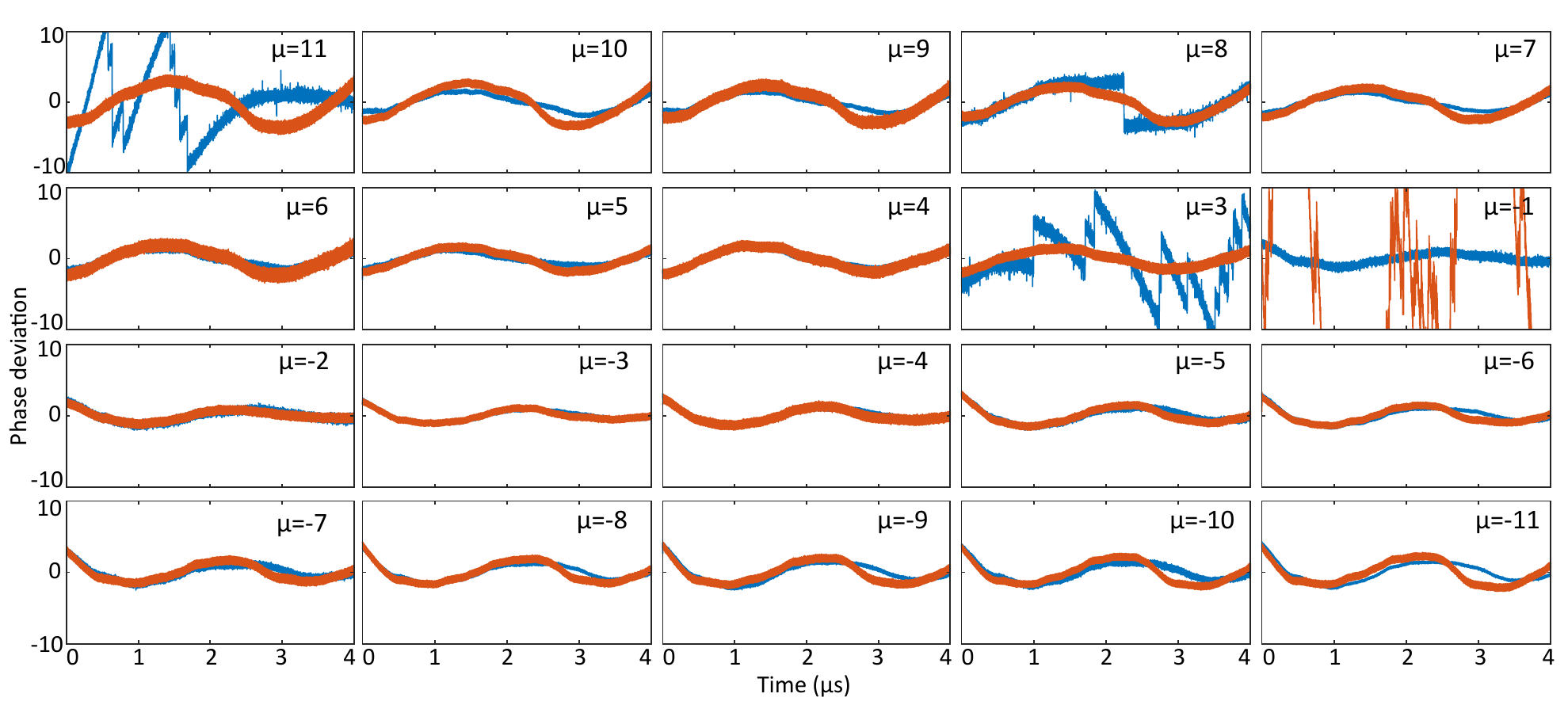}
	\caption{\textbf{Phase deviation of the chirped comb lines.}
	 Blue/Red - Signal/LO comb line. 
	 The phase of the signals was obtained via the Hilbert transform. The subsequent subtraction of quadratic fit (corresponds to linearly changing frequency of the chirp) results in phase deviation.
	}
	\label{fig_SI_2combs_phase}
\end{figure*}

\newpage
\begin{figure*}[!htbp] 
	\includegraphics[width=\linewidth]{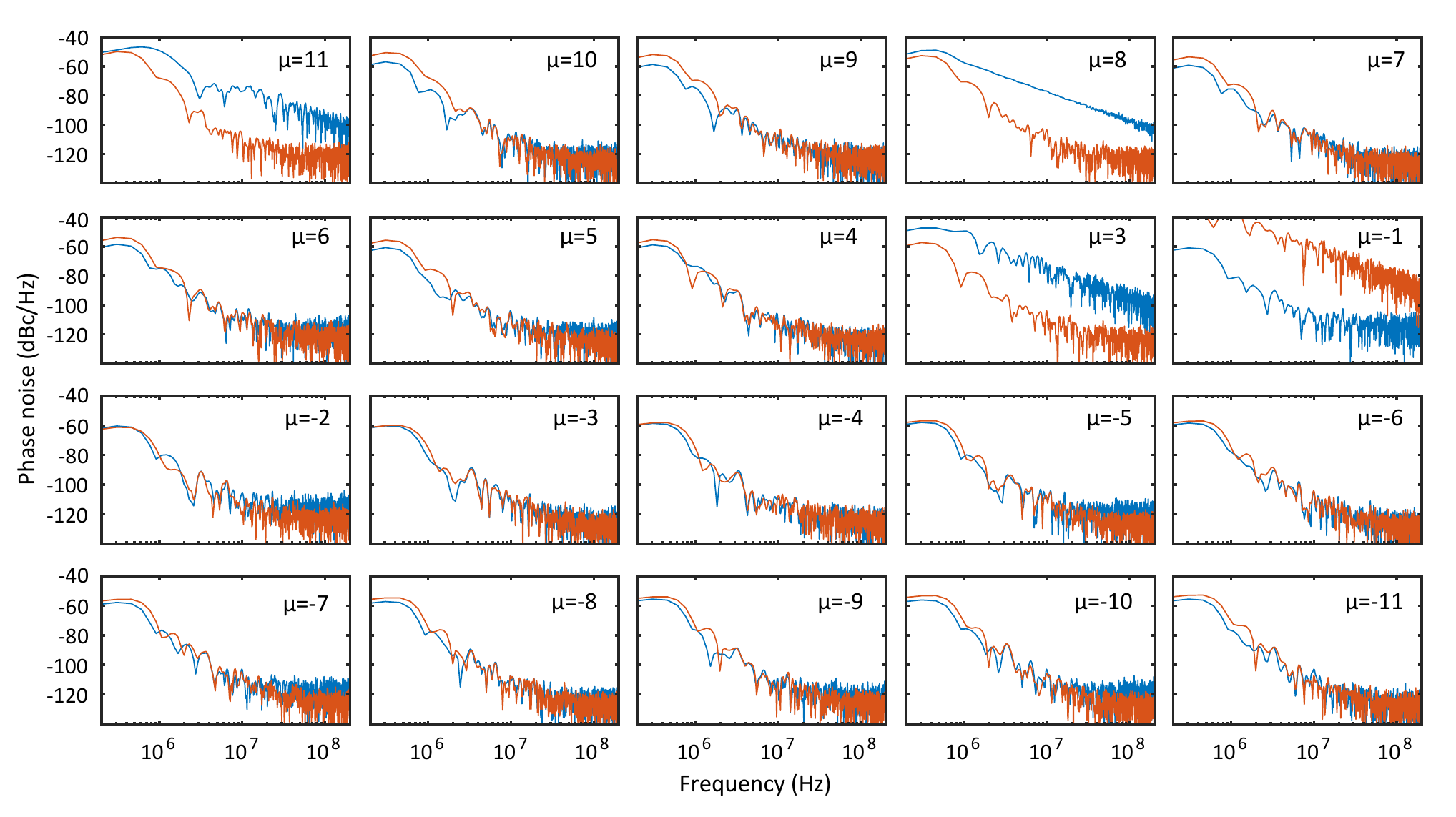}
	\caption{\textbf{Phase noise of Signal and LO combs.}
	Blue/Red - Signal/LO comb line. 
	}
	\label{fig_SI_2combs_psd}
\end{figure*}

\newpage
\begin{figure*}[!htbp] 
	\includegraphics[width=\linewidth]{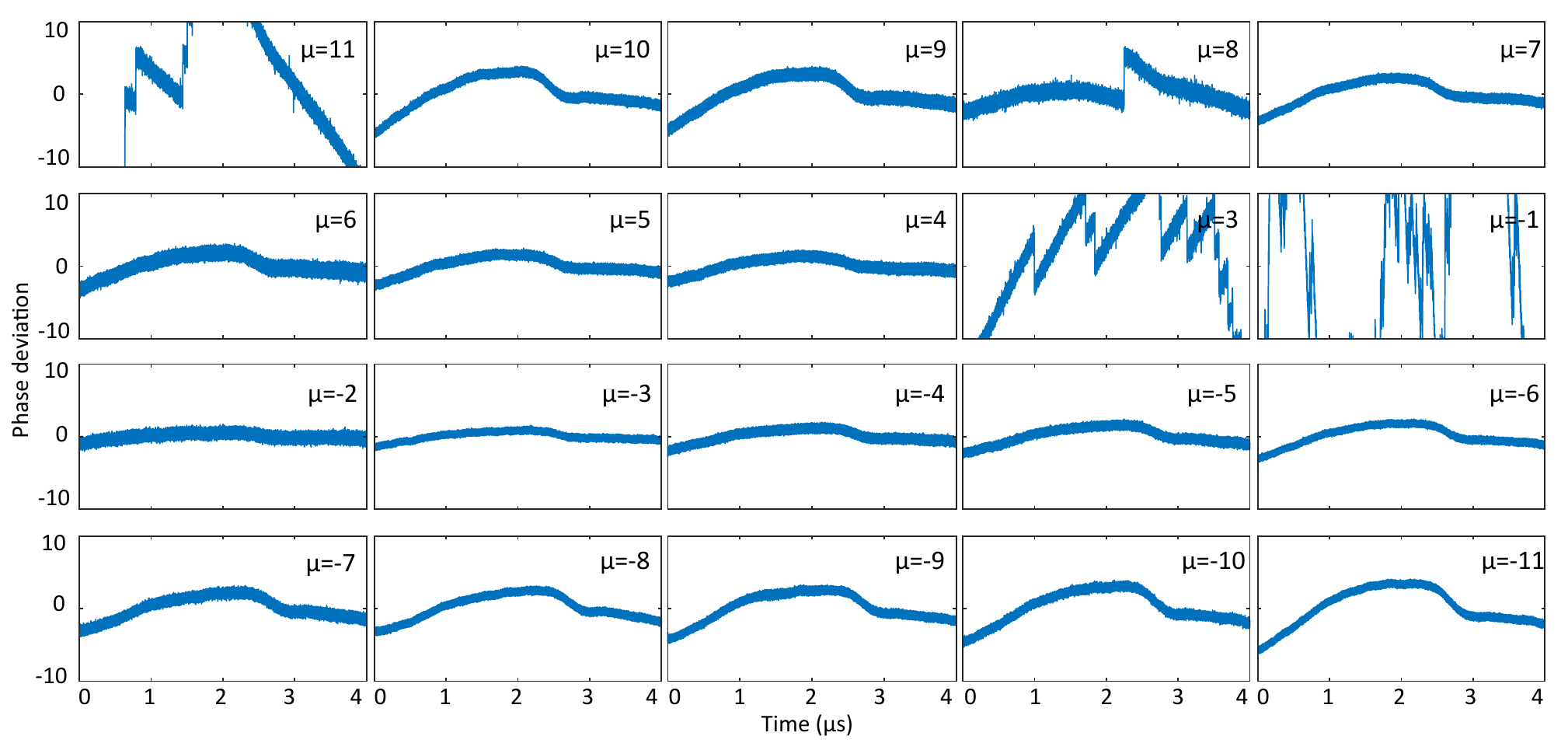}
	\caption{\textbf{Differential phase deviation of the chirped comb lines.}
	The phase deviation is obtained by subtracting the phase of the Signal comb from the phase of the LO comb for a given channel and subsequent subtraction of linear fit.
	}
	\label{fig_SI_rel_phase}
\end{figure*}

\newpage
\begin{figure*}[!htbp] 
	\includegraphics[width=\linewidth]{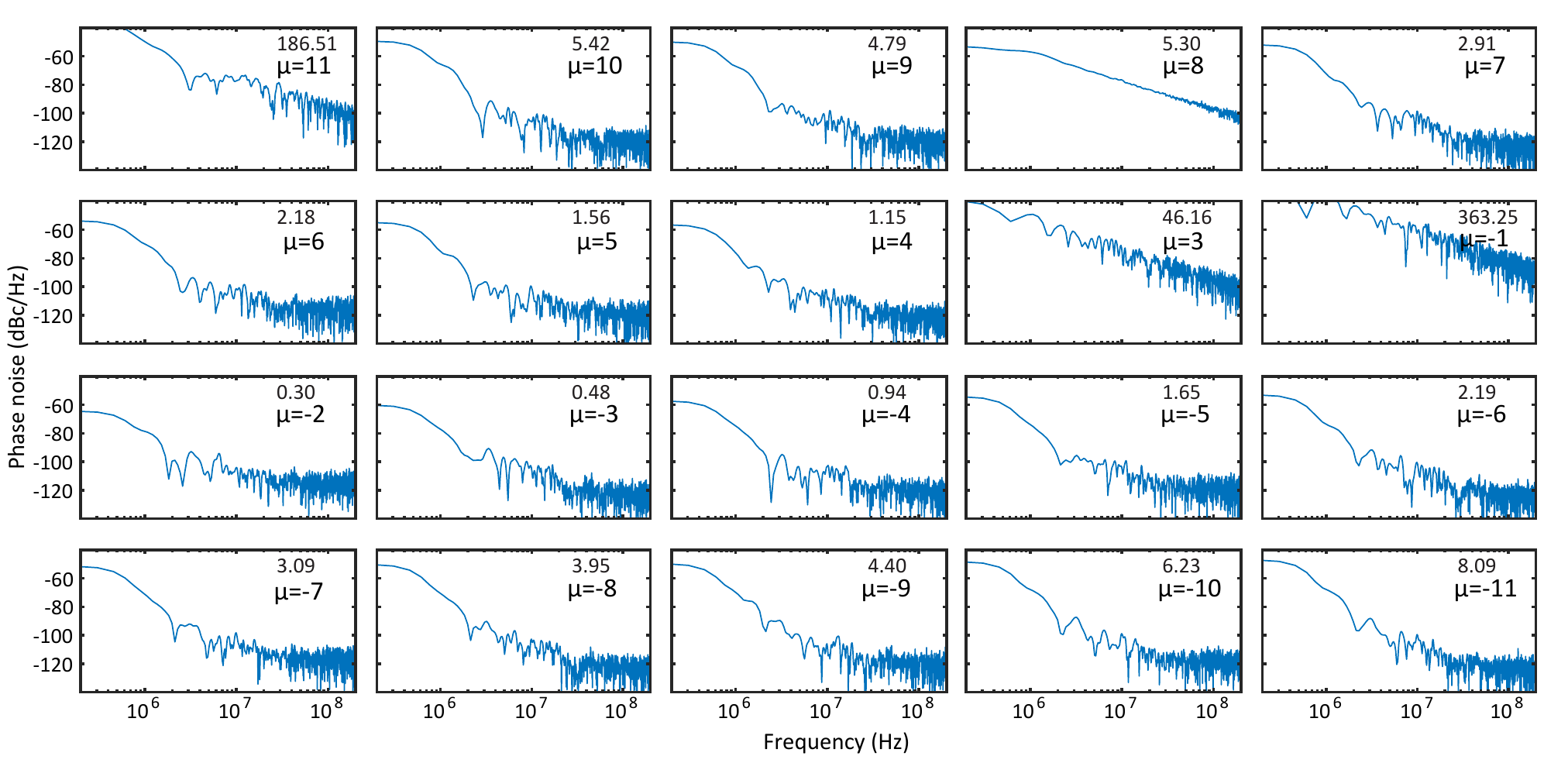}
	\caption{\textbf{Phase noise of the beatnote signal between Signal and LO.}
	The number in the top right of every plot corresponds to the integral of the PSD. Heuristically, the value less than $2/\pi$ corresponds to the Fourier-limited beatnote linewidth.
	}
	\label{fig_SI_rel_psd}
\end{figure*}

\newpage
\begin{figure*}[!htbp] 
	\includegraphics[width=\linewidth]{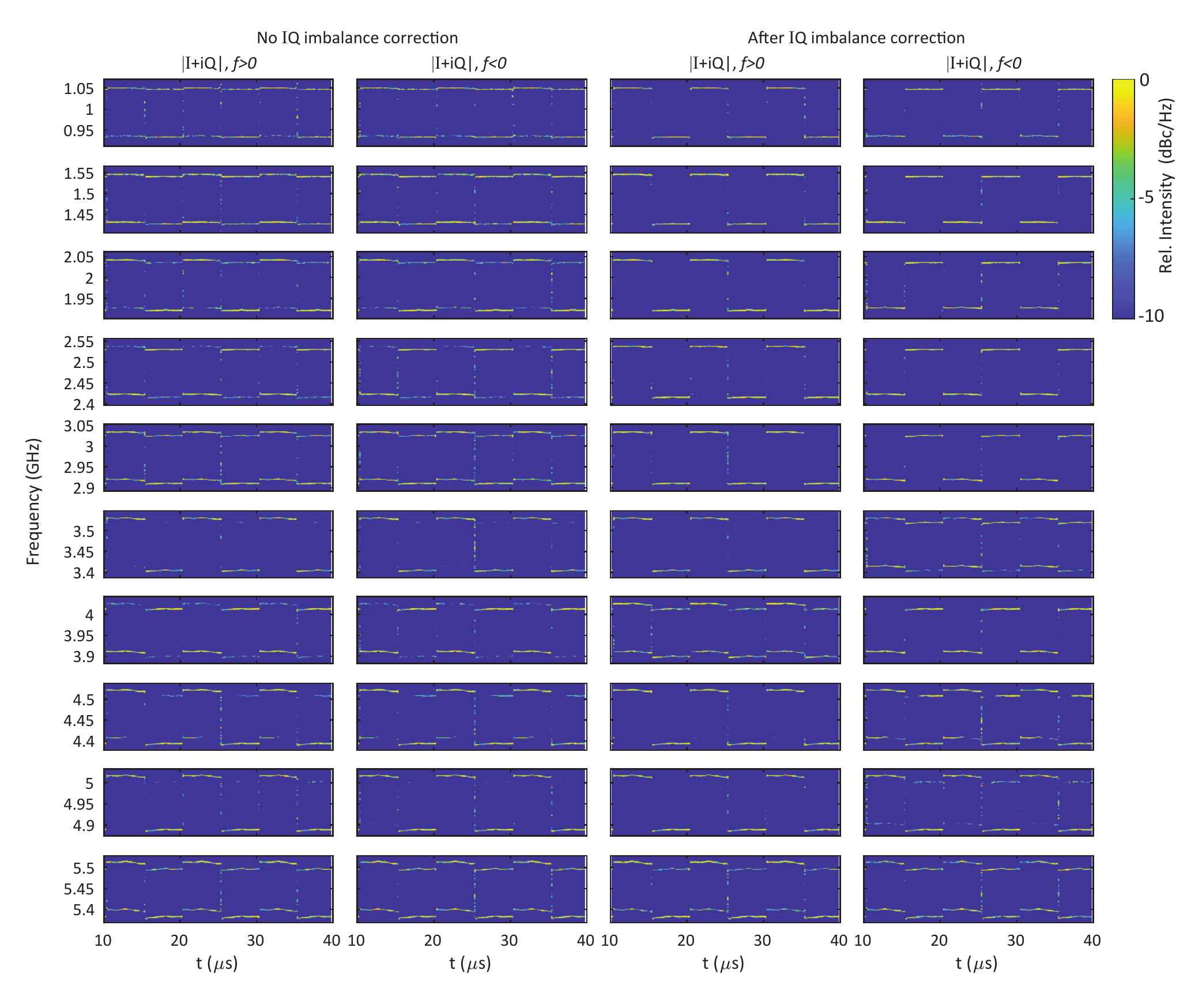}
	\caption{\textbf{IQ imbalance correction.}
	Time-frequency map of delayed heterodyne beat spectroscopy for $\mu = \pm2,\pm11$. First two columns correspond to positive and negative channels without IQ imbalance correction, while second two columns are with IQ imbalance correction. ENBW 2.45~MHz.
	}
	\label{fig_SI_IQ}
\end{figure*}

\newpage
\begin{figure*}[!htbp] 
	\includegraphics[width=\linewidth]{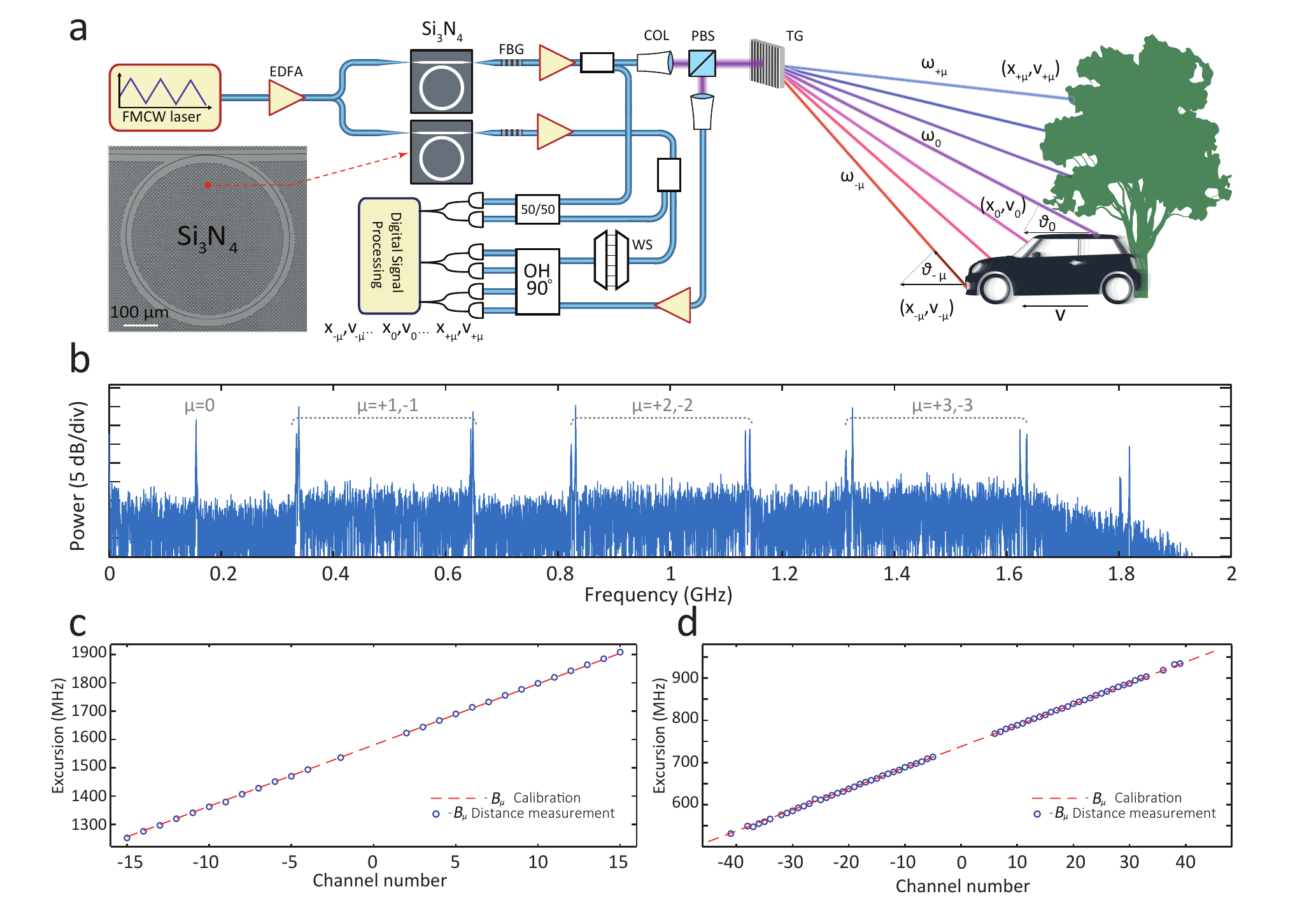}
	\caption{\textbf{Excursion calibration.}
	a)~Architecture of the multiheterodyne parallel FMCW LiDAR. A pump laser with triangular frequency modulation drives two distinct optical microresonators with slightly different radii, which serve as signal and LO in the experiment. 
	After pump filtering (FBG) and amplification (EDFA), the signal comb is spatially dispersed over the target area using diffractive optics. Each signal comb tooth $\mu$ represents an independent FMCW ranging channel measureing distance $x_{\mu}$ and velocity  $v_{\mu}$. 
	All channels are simultaneously superimposed with the amplified LO comb on a coherent receiver. Waveshaper (WS) is used filter out excess ASE noise around the pump of the LO.
	Extra reference balanced photo-diode is used to measure frequency excursion.
	b)~Interferogram of a LO and signal combs beating recorded over one period ($10 \mu s$) on the reference photodiode with 1.6GHz bandwidth. Beatings of 6 channels ($\mu = \pm1,\pm3$) allow $\Delta f_\mathrm{rep}$ and $B_{\mu}$ to be retrieved and used for the processing of the measured data.
	c)~Two frequency excursion measurements with 100 GHz chips. Red dashed line is an extrapolation of excursion obtained from the reference delayed heterodyne measurement. Blue circles correspond to an excursion obtained from a lidar distance measurement of a carton block, where all the distances are the same for distinct channels. 
	d)~Same as c), but with 35 GHz chips
	}
	\label{fig_SI_calibration}
\end{figure*}

\newpage
\begin{figure*}[!htbp] 
	\includegraphics[width=\linewidth]{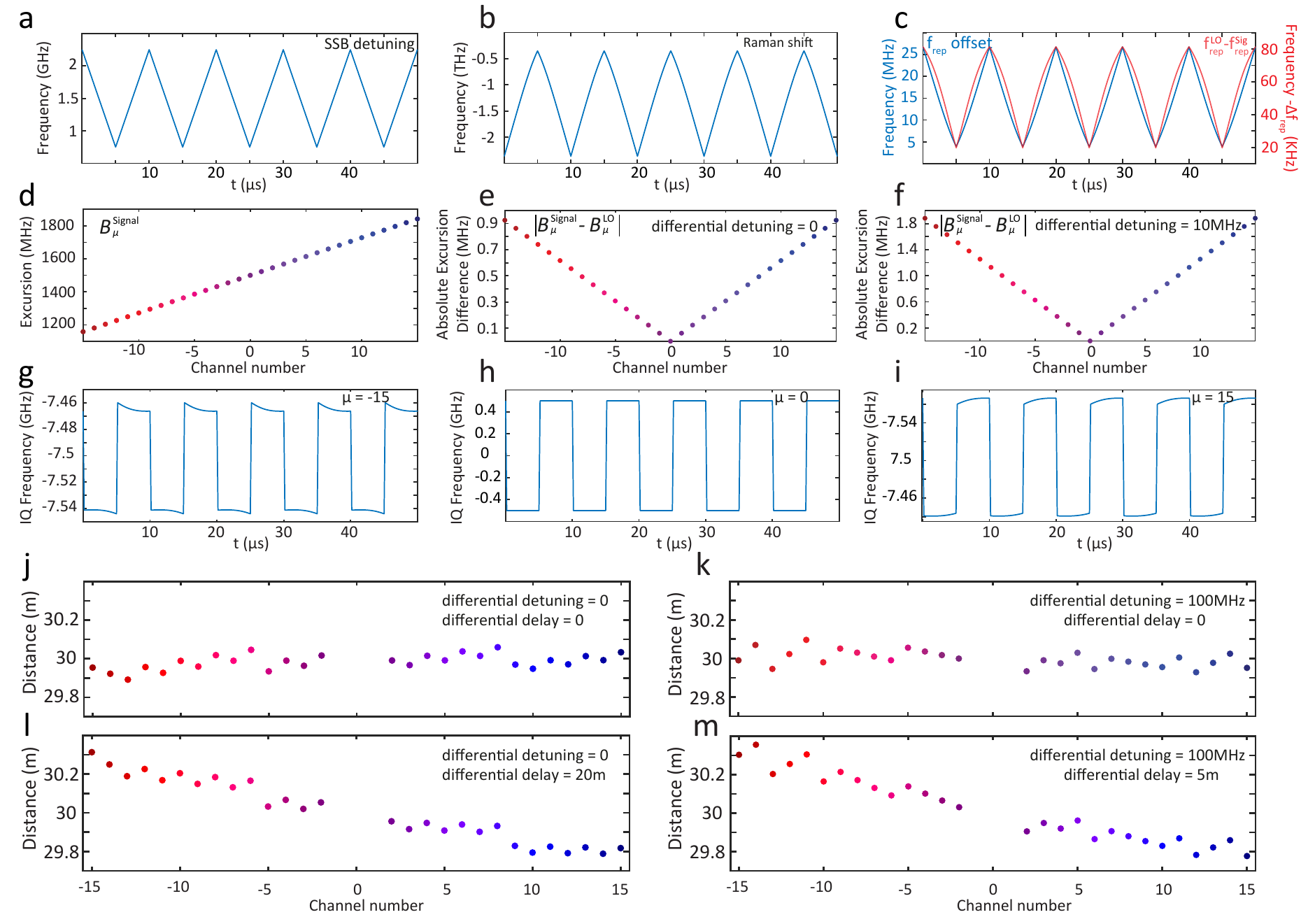}
	\caption{\textbf{Raman nonlinearity impact in 100 GHz resonators.}
	a)~Simulation of a pump cavity detuning representing a triangular ramp.
	b)~Simulation of a Raman shift experienced by a Soliton while being frequency swept.
	c)~Blue - $f_\mathrm{rep}$ offset from a 100GHz caused by Raman nonlinearities. Red - $\Delta f_\mathrm{rep}$ offset from a 500 MHz (100 and 100.5 GHz free spectral ranges were used in this simulation) for the two solitons while being frequency swept due to Raman nonlinearities.
	d)~Simulated frequency excursion for 100 and 100.5 GHz resonators due to Raman effect.
	e,f)~Simulated absolute frequency excursion difference considering soliton-cavity detuning to be the same for two resonators and have 10 MHz difference correspondingly.
	g,h,i) ~Simulated time-frequency maps obtained from a beating of signal and LO combs. Difference in excursion due to Raman effect results in deformed trapezoidal traces. Channels $\mu = -15,0,+15$ correspondingly.
	j,k,l,m)~Simulated distance measurements calculated from time-frequency maps above. The term differential delay stands for the difference between optical delay of signal and LO for the real measurement on one side and optical delay of signal and LO for the reference calibration measurement on the other side.	
	}
	\label{fig_SI_Raman}
\end{figure*}

\newpage
\begin{figure*}[!htbp] 
	\includegraphics[width=0.6\linewidth]{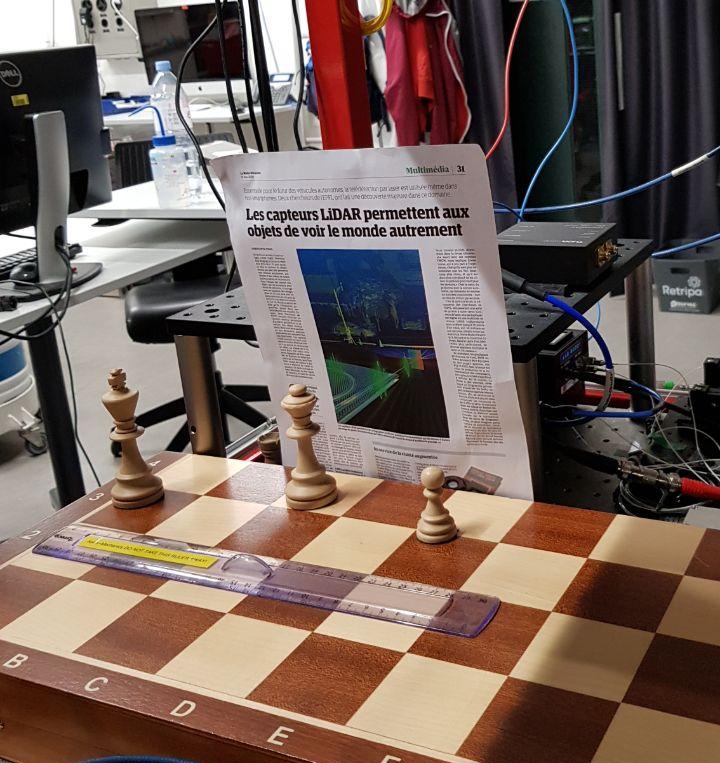}
	\caption{\textbf{Photograph of the imaged scene.}
	~Pawn, Queen and King chess figures used for the lidar distance measurement.}
	\label{fig_SI_chess}
\end{figure*}
 
\begin{figure*}[!htbp] 
	\includegraphics[width=\linewidth]{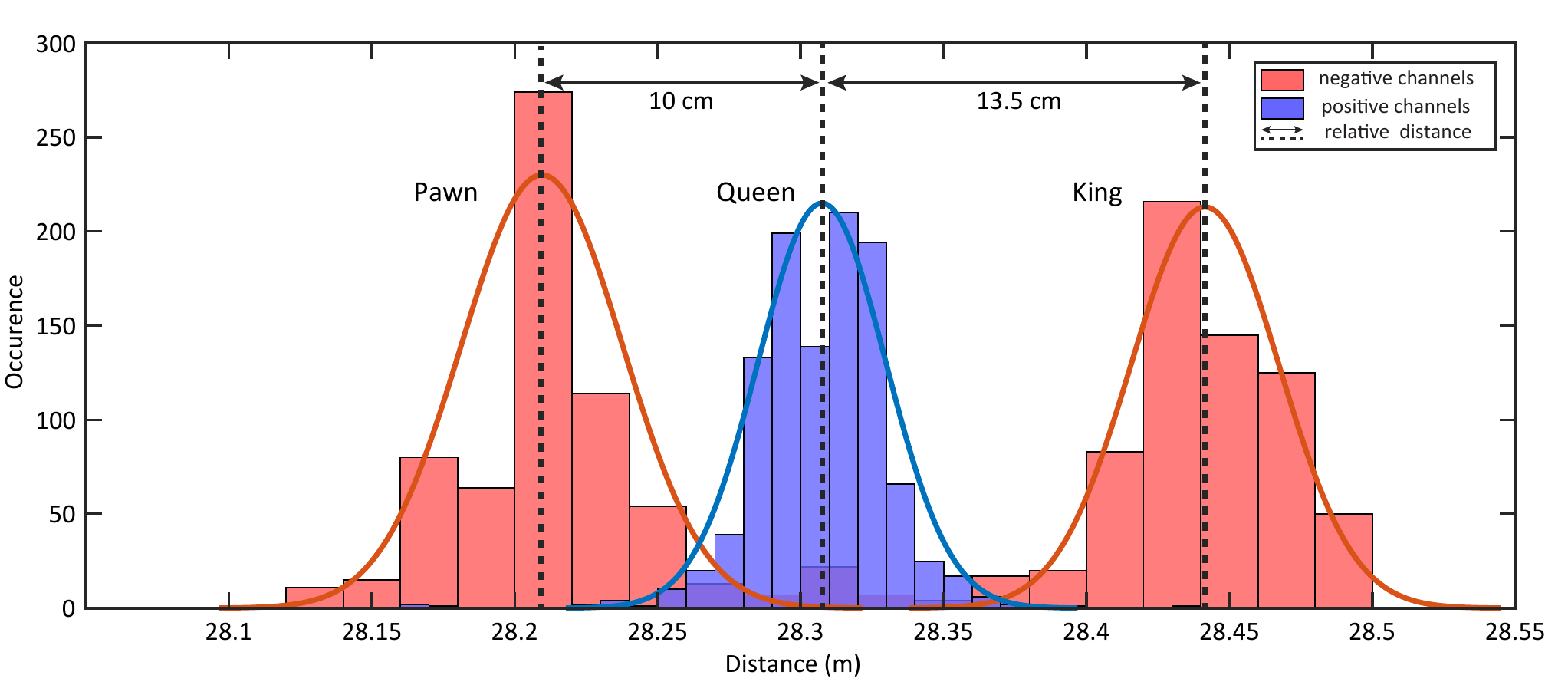}
	\caption{\textbf{Chess ranging measurement relative accuracy.}
	~Measured relative distance for chess figures ranging experiment. Red histogram corresponds to the pixels in negative channels that illuminated Pawn and King, while blue histogram corresponds to Queen illuminated by positive channels. Histograms depict 136*28 measured pixels, where one line of 28 pixels was acquired during 10~$\mu$s time.} 
	\label{fig_SI_uncertainty}
\end{figure*}

\newpage
\begin{figure*}[!htbp] 
	\includegraphics[width=\linewidth]{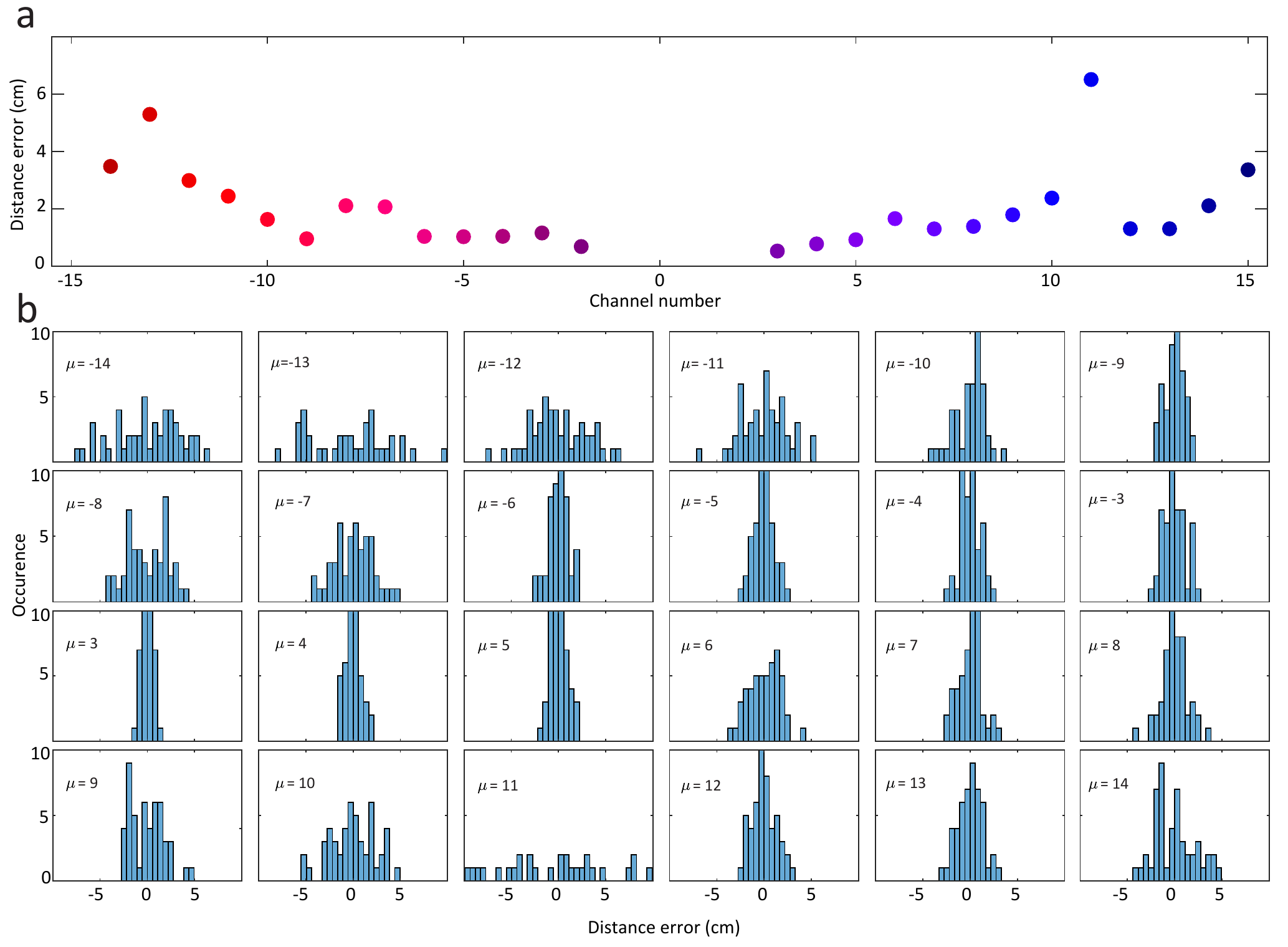}
	\caption{\textbf{Distance precision of the Dual-Comb LiDAR.}
	a)~Channel-dependent distance precision.
	b)~Histograms obtained during 49 distance measurements of the static object.}
	\label{fig_SI_precision}
\end{figure*}

\newpage
\begin{figure*}[!htbp] 
	\includegraphics[width=\linewidth]{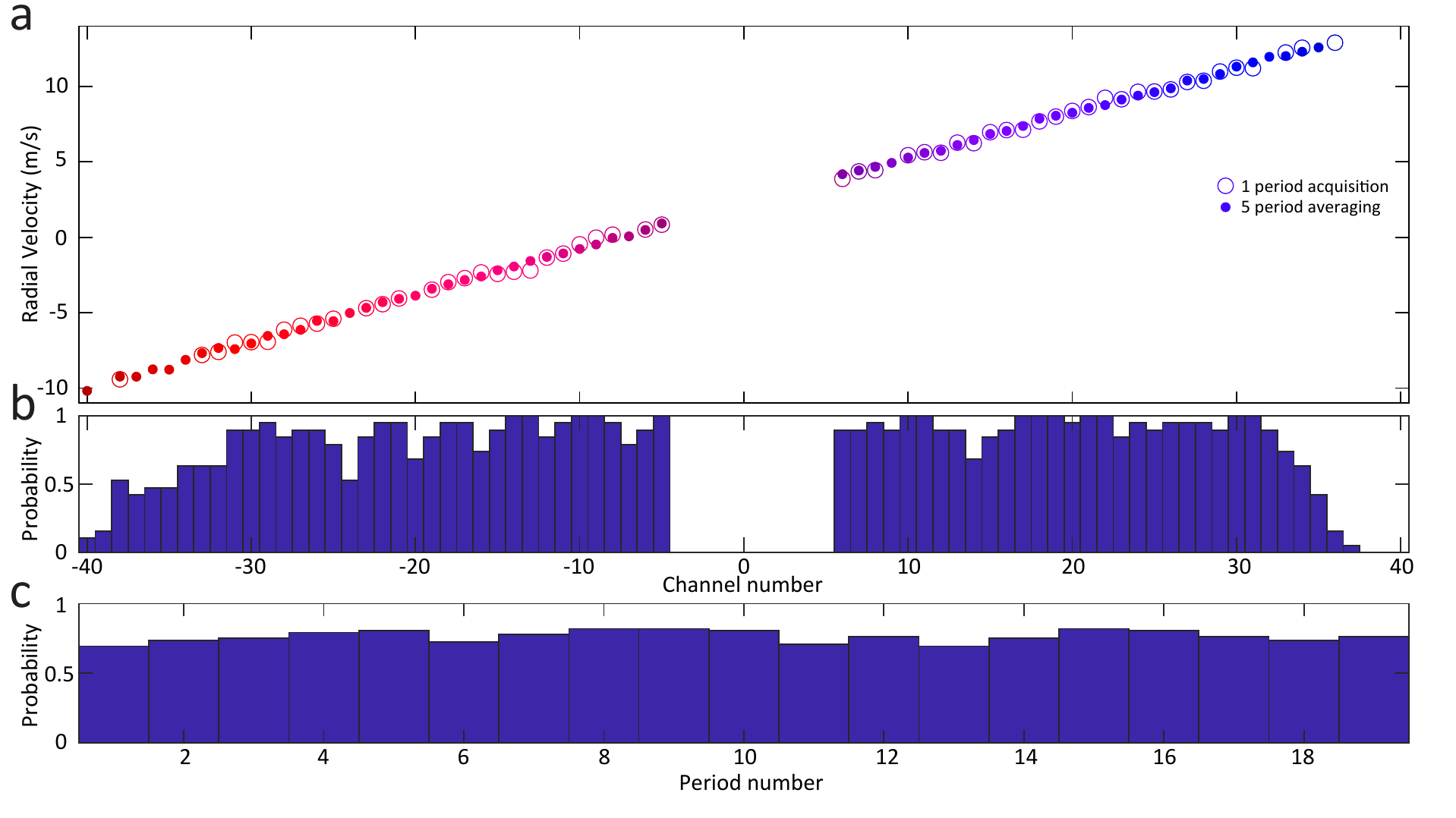}
	\caption{\textbf{Velocimetry measurements statistics.}
	a)~Multichannel velocity measurement for the flywheel rotating at around 162~Hz for a single 10~$\mu$s scan (open circles) and five frame stacking (filled circles).
	b)~Probability of a pixel to be detected in channel $\mu$. Statistics are obtained over 19 periods measurement time ($190 \mu s$). The roll-off at large distance from the pump laser originates from the limited optical amplifier bandwidth.
	c)~Probability of detection over a given time frame ($10 \mu s$). The probability is calculated as number of total pixels detected in that time frame divided by 81 ($\mu = -40,+40$)).
	}
	\label{fig_SI_velocity_statistics}
\end{figure*}

\end{document}